%% file: 0_main.tex
\documentclass[10pt,journal,compsoc]{IEEEtran}

\newcommand{\ie}{\emph{i.e., }}
\newcommand{\eg}{\emph{e.g., }}

\newcommand{\wrt}{\emph{w.r.t. }}

\newcommand{\aka}{\emph{a.k.a. }}

\usepackage{enumitem}
\usepackage{verbatim}
\usepackage[ruled]{algorithm2e}
\usepackage{multirow}
\usepackage{subfigure} 
\usepackage{ragged2e}
\usepackage{caption}
\usepackage{hyperref}
\usepackage{amsmath}
\usepackage{cleveref}
\usepackage{graphicx}
\usepackage[normalem]{ulem}
\useunder{\uline}{\ul}{}
\usepackage{amsthm}
\usepackage{bm}
\usepackage{setspace}
\usepackage{diagbox}

\begin{document}
\title{Causal Intervention for Fairness in Multi-behavior Recommendation}

\author{Xi Wang$^1$, Wenjie Wang$^{2*}$, Fuli Feng$^{3*}$, Wenge Rong$^{4}$, Chuantao Yin$^5$, Zhang Xiong$^4$\\
\IEEEauthorblockA{$^1$ State Key Laboratory of Complex \& Critical Software Environment, Beihang University, China}\\
\IEEEauthorblockA{$^2$ National University of Singapore, Singapore}\\
\IEEEauthorblockA{$^3$ University of Science and Technology of China, Hefei, China}\\
\IEEEauthorblockA{$^4$ School of Computer Science, Beihang University, Beijing, China}\\
\IEEEauthorblockA{$^5$ Sino-French Engineer School, Beihang University, Beijing, China}\\ 
 Email: 
 samuel.wang.xi@centralepekin.cn,wangwenjie@u.nus.edu, fulifeng93@gmail.com,\{w.rong, chuantao.yin, xiongz\}@buaa.edu.cn
\thanks{* Corresponding author. }
 }

\IEEEtitleabstractindextext{%
\begin{abstract}
Recommender systems usually learn user interests from various user behaviors, including clicks and post-click behaviors (\eg like and favorite, which reflects the true interests of users~\cite{ma2018entire}). However, these behaviors inevitably exhibit popularity bias, leading to some unfairness issues: 1) for items with similar quality, more popular ones get more exposure; and 2) even worse the popular items with lower popularity might receive more exposure. Existing work on mitigating popularity bias blindly eliminates the bias and usually ignores the effect of item quality. We argue that the relationships between different user behaviors (\eg conversion rate) actually reflect the item quality. Therefore, to handle the unfairness issues, we propose to mitigate the popularity bias by considering multiple user behaviors.

In this work, we examine causal relationships behind the interaction generation procedure in multi-behavior recommendation. Specifically, we find that: 1) item popularity is a confounder between the exposed items and users' post-click interactions, leading to the exposure unfairness; and 2) some hidden confounders (\eg the reputation of item producers) affect both item popularity and quality, resulting in the quality unfairness. To alleviate these confounding issues, we propose a causal framework to estimate the causal effect, which leverages \textit{backdoor adjustment}~\cite{pearl2009causality} to block the backdoor paths caused by the confounders. In the inference stage, we remove the negative effect of popularity and utilize the good effect of quality for recommendation. 
Experiments on two real-world datasets validate the effectiveness of our proposed framework, which enhances fairness without sacrificing recommendation accuracy on post-click behavior.
\end{abstract}
\vspace{-10pt}


}



\maketitle
\input{1_intro}
\input{2_method}
\input{4_experiments}

\input{5_related}

\input{6_conlusion}
\input{8_acknowledgement}
\bibliographystyle{IEEEtran}
\bibliography{0_main}
%

\end{document}

%% file: 1_intro.tex
\section{Introduction} \label{sec:intro}

Recommender systems are becoming increasingly important for personalized information filtering on various platforms, such as E-commerce and video-streaming. 
To learn user preferences, a promising choice is jointly mining multiple user behaviors.
Specifically, user behaviors can be divided into click\footnote{For brevity, we use clicks to represent various implicit feedback, such as page-view, listen, play, and download.} and post-click behaviors (\eg buy and like)~\cite{ma2018entire}.
Due to the uneven distribution of user behaviors, recommender models typically suffer from popularity bias, which hinders the model fidelity~\cite{zhang2021causal}. 
Worse still, recommender systems tend to amplify popularity bias, \eg over-recommending popular items, causing various issues such as unfairness or filter bubbles~\cite{Calibrated}.
Therefore, this work focuses on mitigating the impact of popularity bias from the perspective of multiple user behaviors.

Existing work on alleviating popularity bias in recommendation mainly falls into two categories:
\begin{itemize}[leftmargin=*]
    \item \textit{Causal inference for debiasing} inspects the causal relationships behind interaction generation procedure, and removes the confounding effect of popularity through causal intervention~\cite{zhang2021causal,wang2021deconfounded} or counterfactual inference~\cite{gupta2021causer}.
    However, they focus on reducing the impact of popularity bias or using popularity to achieve better accuracy, but do not pay attention to the impact of popularity on fairness in recommendation.
    Besides, they ignore the relationships between different behaviors and only mitigate popularity bias on one of them, which might lead to harmful unfairness issues.
    \item \textit{Multi-behavior recommendation} uses multi-task learning to predict both click and post-click behaviors, which usually incorporates inverse propensity scoring (IPS) to reduce the impact of popularity on click behaviors~\cite{zhang2020large, ma2018entire, wen2020entire}. Nevertheless, the propensity is hard to estimate properly, and thus IPS-based methods often suffer from the high variance problem~\cite{saito2020unbiased}.
\end{itemize}

\begin{figure}
\centering
\subfigure[Average exposure and ranks of items with similar quality.]
{\label{fig:a}\includegraphics[width=0.23\textwidth]{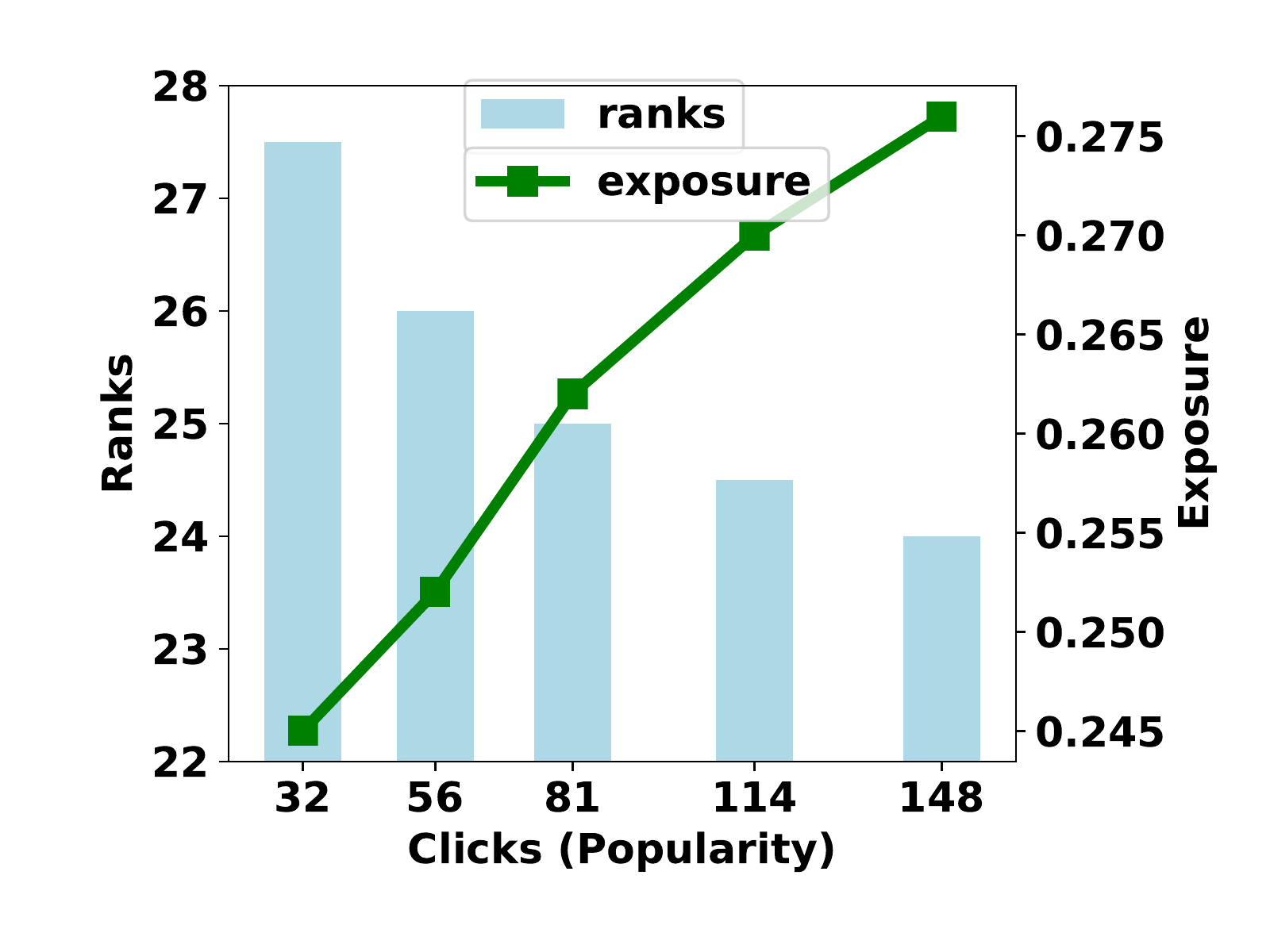}}
\subfigure[Average exposure and ranks of items with same likes.]
{\label{fig:b}\includegraphics[width=0.23\textwidth]{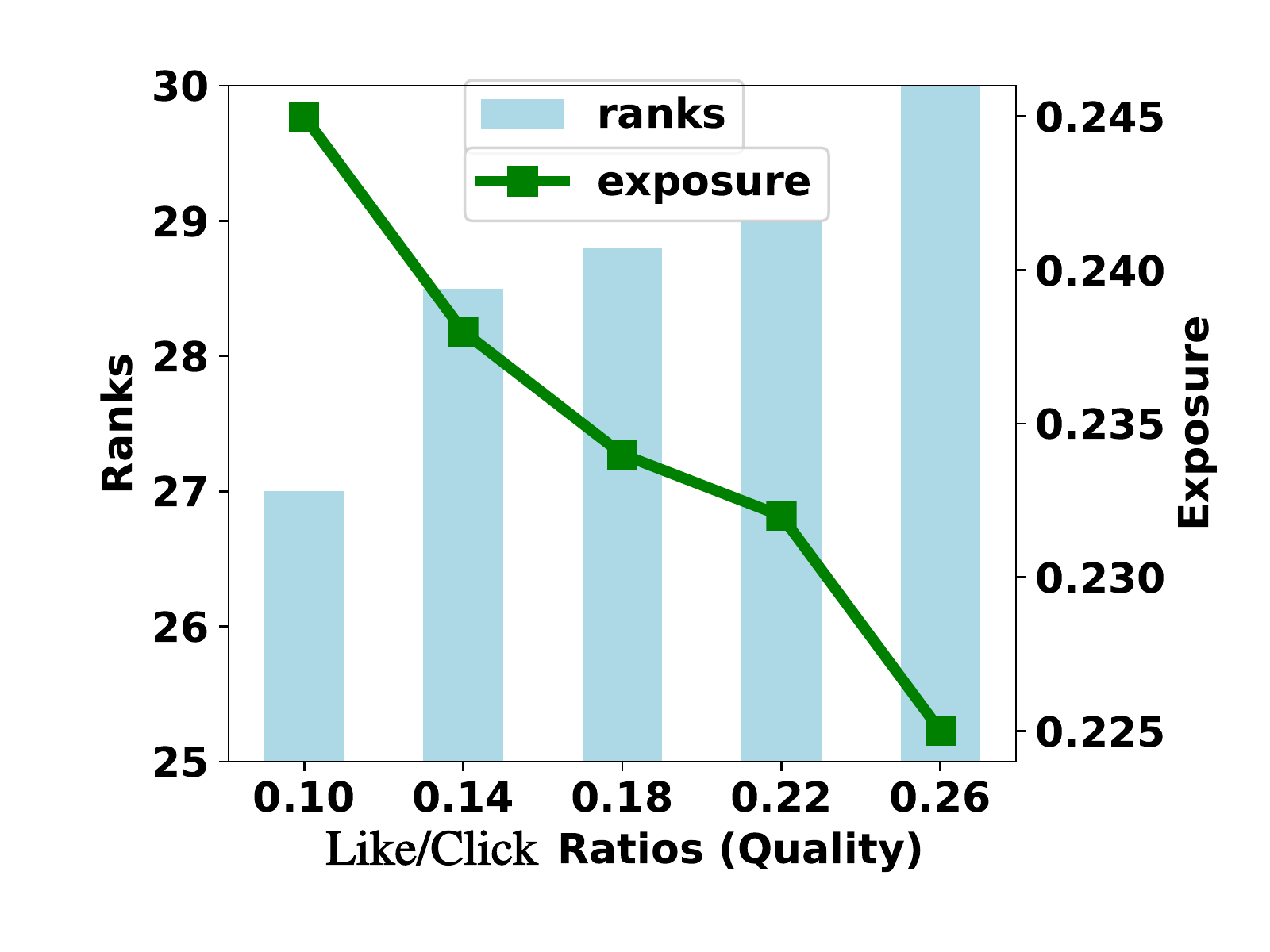}}
\vspace{-10pt}
\caption{Illustration of two kinds of unfairness. (a) Items' ranks and exposure \wrt increasing clicks (popularity) where the item quality is similar according to the same like/click ratio $[0.1, 0.14]$; (b) the ranks and exposure \wrt ascendant like/click ratios where items are selected by the same like number (\ie 5). The results are obtained by training ESMM~\cite{ma2018entire} on the Kwai dataset~\cite{zhang2021causal}.}
\label{fig:rec_unfairnesses}
\vspace{-15pt}
\end{figure}

In this work, we identify two kinds of unfairness caused by popularity bias in multi-behavior recommendation.
Figure~\ref{fig:rec_unfairnesses} provides an empirical evidence with a representative model ESMM~\cite{ma2018entire} on the Kwai dataset~\cite{zhang2021causal}.
1) As shown in Figure \ref{fig:rec_unfairnesses}(a), more popular items will get higher ranks and more exposure opportunities, while having similar quality; and 2) items with lower quality but high popularity might receive more exposure. For example, for items that receive similar post-click behaviors (\eg the same number of likes), the ones with more clicks have smaller like/click ratios and usually have lower quality. However, as illustrated in Figure \ref{fig:rec_unfairnesses}(b), high popularity leads to more exposures on such items. 
To address these unfairness issues, it is critical to mitigating the popularity bias of both click and post-click behaviors.

\begin{figure}[]
\centering
\subfigure[Causal graph of user-item interaction procedure.]{\label{fig:a}\includegraphics[width=0.18\textwidth]{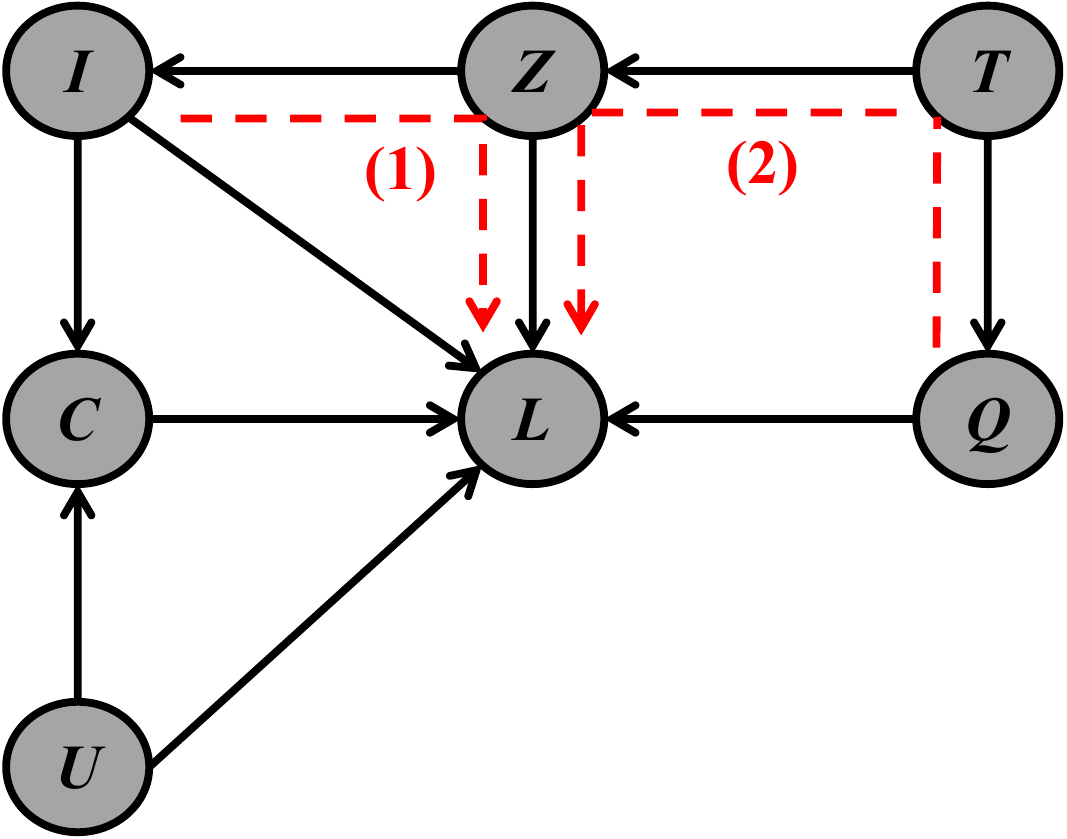}}
\subfigure[Cutting off two backdoor paths by two interventions.]{\label{fig:b} \includegraphics[width=0.21\textwidth]{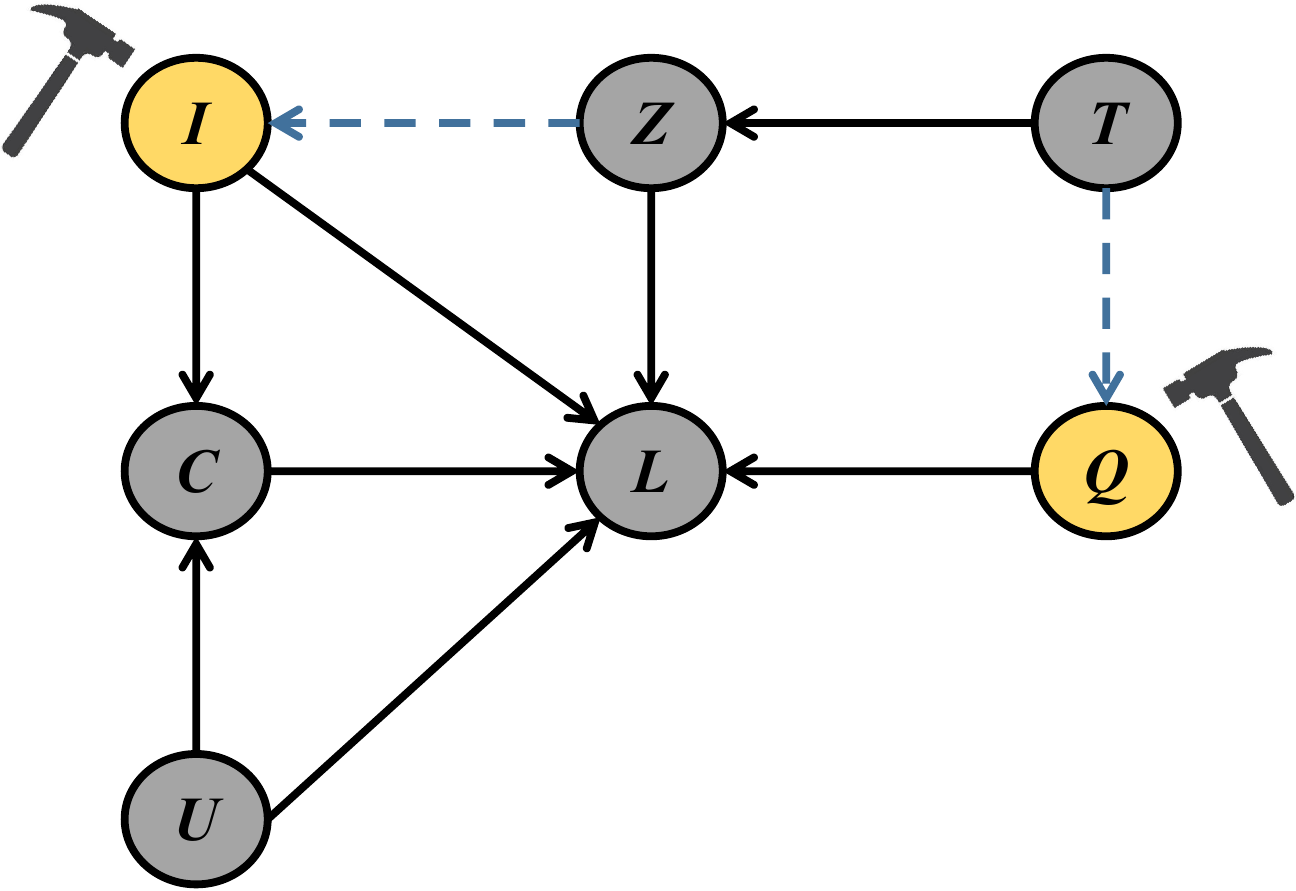}}
\vspace{-10pt}
\caption{Causal graph for multi-behavior recommendation. (1) and (2) are two backdoor paths. The confounding effects of $Z$ from $I$ to $L$ and from $Q$ to $L$ are eliminated by the two interventions on $I$ and $Q$.}
\vspace{-10pt}
\label{fig:rec_graph}
\end{figure}

To explore the reasons for the unfairness in multi-behavior recommendation, we inspect the causal relationships by the causal graph in Figure \ref{fig:rec_graph}: 
$U$ and $I$ represent user and item, $C$ and $L$ denote the user's click probability and post-click probability on the item, $Z$ is item popularity, $Q$ is item quality and $T$ represents some unobserved variables. Firstly, we will introduce the definition of Backdoor Criterion~\cite{pearl2009causality} in causal intervention. In order to measure the direct effect that one variable, say X, has on another one, say Y, we must first make sure to isolate the effect from any other spurious correlations that might be present. The easiest way to do this is to make sure all non-causal paths between X and Y are blocked off. We can easily identify the variables we need to condition on by applying the so called ‘backdoor criterion’ which is defined as: Given an ordered pair of variables (X, Y) in a directed acyclic graph G, a set of variables Z satisfies the backdoor criterion relative to (X, Y) if no node in Z is a descendant of X, and Z blocks every path between X and Y that contains an arrow into X. And Z is a confounder between variables X and Y.

In the proposed causal graph, $Z\rightarrow L$ means that popularity affects user's post-behaviors because of the user's herd mentality (\aka conformity),
and $Z\rightarrow I$ denotes the impact of popularity on item exposure: popular items are more likely to be recommended by the systems~\cite{Mattheweffect, zhang2021causal}. As such, $Z$ is a confounder, causing the backdoor path $I\leftarrow Z\rightarrow L$ from $I$ to $L$. Consequently, for items with similar quality, the more popular ones will receive higher post-click probability $L$ due to the confounding effect of $Z$. 
As to the quality unfairness, we find that some unobserved confounders $T$ (\eg reputation of item producer) affect both item popularity $Z$ and quality $Q$, resulting in another backdoor path $Q\leftarrow T\rightarrow Z \rightarrow L$ from $Q$ to $L$ (Refer to Section \ref{ssec:causal graph} for more details about the causal graph). As a result, items with the worse quality might correlate with high popularity because of unobserved $T$, and thus they are more frequently recommended. 

The key to solving the two unfairness issues lies in cutting off the two backdoor paths: $I\leftarrow Z\rightarrow L$ and $Q\leftarrow T\rightarrow Z \rightarrow L$. 
To block the first backdoor path, we can incorporate \textit{backdoor adjustment}~\cite{pearl2009causality} to perform the intervention $do(I=i)$ thanks to the observed $Z$~\cite{pearl2009causality, zhang2021causal}. 
However, $T$ and $Q$ are usually unobserved variables in the second backdoor path $Q\leftarrow T\rightarrow Z \rightarrow L$, making backdoor adjustment intractable. Fortunately, the items' post-click conversion ratio (\ie the number of post-click behaviors divided by the click number) is able to reflect the item quality (See Figure \ref{fig:quality_zhihu} for the evidence), and thus we use it as the approximation of $Q$~\cite{lalmas2015promoting}. Furthermore, to cut off the second backdoor path via backdoor adjustment, we can control $Z$ without accessing the values of $T$ because $Z$ blocks all the backdoor paths from $Q$ to $L$~\cite{pearl2009causality}.
As such, we propose a \textbf{M}ulti-\textbf{B}ehavior \textbf{D}ebiasing framework (MBD) to block the two backdoor paths by two interventions. 
Specifically, MBD additionally considers item quality for prediction and estimates the post-click probability $P (L|U, do(I), do(Q))$ during training. Thereafter, MBD removes the bad effect of $Z$ on $L$ and keeps the good effect of $Q$ for inference. 
We conduct experiments on two real-world datasets, showing that MBD can effectively alleviate the two unfairness issues and achieve superior accuracy. We release the code and data at \url{https://anonymous.4open.science/r/MBD-DCBC/}.

The main contributions of our work are as follows:
\begin{itemize}[leftmargin=*]
    \item We study a new problem of popularity bias in multi-behavior recommendation, and address the two consequent unfairness issues from a causal view.
    \item We propose a multi-behavior debiasing framework, which blocks the backdoor paths leading to two kinds of unfairness via two interventions.
    \item To achieve the causal intervention in the experimental model, during the training process, we model the influence of popularity on the post-click recommendation score, which contains the backdoor path impact between Z and L. To remove the influence of popularity on item exposure and quality of recommendations by cutting off the backdoor path, we set the popularity as a fixed constant in the inference process.
    \item Extensive experiments on two real-world datasets verify the effectiveness of MBD in mitigating the unfairness issues and maintaining the recommendation accuracy.
\end{itemize}

%% file: 2_method.tex
\vspace{-5pt}
\section{Methodology}

In this section, we first formulate two fairness problems by considering multiple user behaviors, and then we discover the reasons for unfairness from a causal view. Lastly, we introduce the MBD framework to mitigate the unfairness problem via two interventions.
\vspace{-5pt}
\subsection{Fairness in Multi-behavior Recommendation}\label{sec:fairness_formulation}





Previous work either removes popularity bias (\eg MACR~\cite{wei2021model}) or leverages popularity via causal inference to achieve better accuracy (\eg PDA~\cite{zhang2021causal}). 
However, they only focus on the accuracy regarding one kind of user behavior, ignoring the relations between multiple user behaviors and the unfairness caused by popularity bias.
Different from MACR~\cite{wei2021model} and PDA~\cite{zhang2021causal}, we propose utilizing multiple user behaviors to measure item quality, and mitigate the unfairness issues by explicitly modeling the relations between popularity and quality. In this paper, we focus on the relations between popularity and item quality in multi-behavior recommendation, and then find two kinds of unfairness problems. Next, we will introduce the definition and the quantitative calculation of two kinds of fairness.



\vspace{5pt}

\begin{figure}[t]
    \centering
    \includegraphics[width=.25\textwidth]{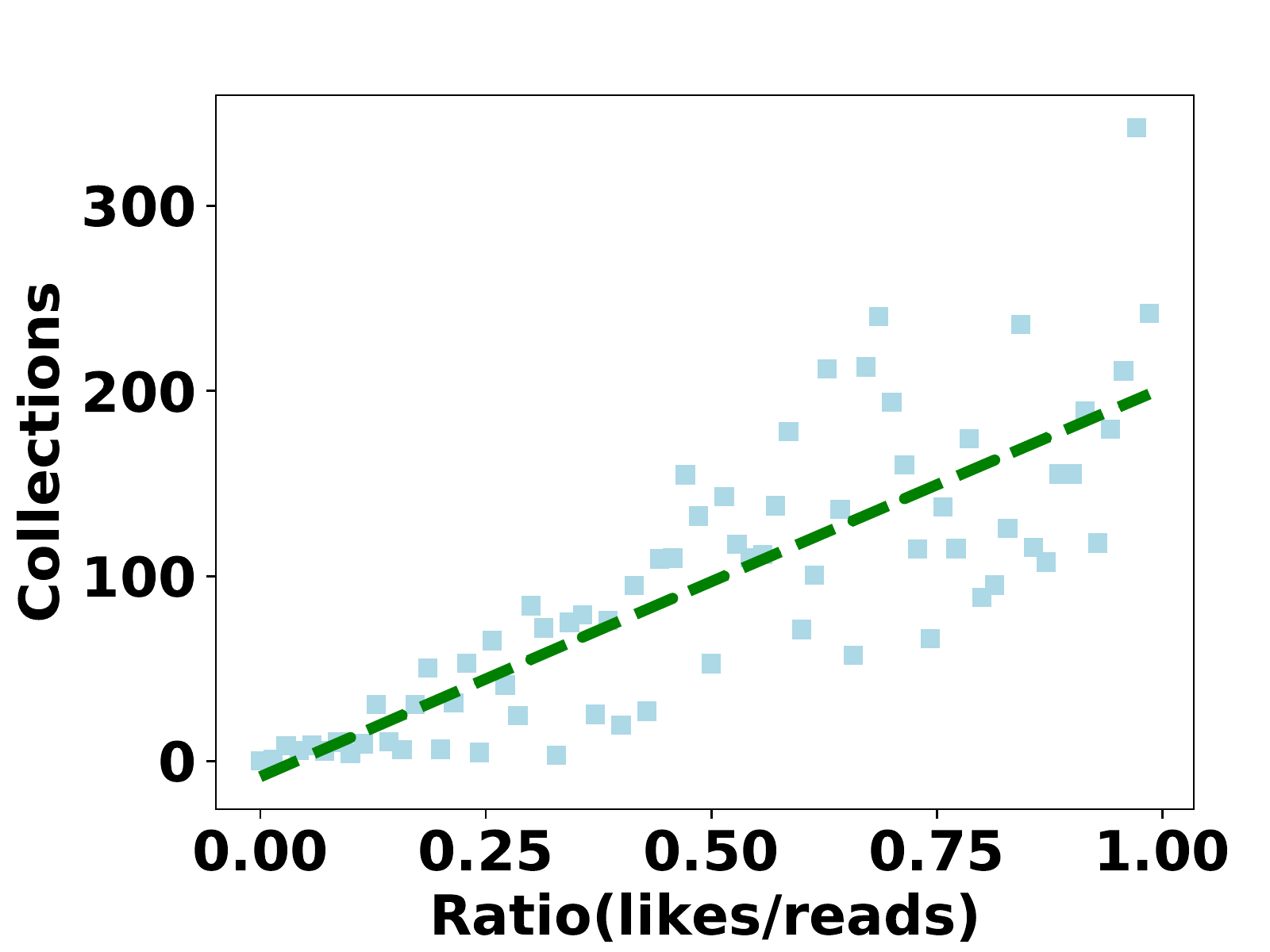} 
    \vspace{-5pt}
    \caption{Collections of articles \wrt post-click conversion ratio from reads to likes.}
    \label{fig:quality_zhihu}
    \vspace{-15pt}
\end{figure}

\noindent\textbf{$\bullet$ Approximation of item quality.} Since not all the datasets have the item quality score, it is difficult to estimate it directly. In this work, we propose to use the item's post-click conversion ratio to make an approximation. 
Previous research~\cite{lalmas2015promoting} and our experiments both demonstrate the positive correlation between the item quality and post-click conversion ratio. 
We conduct experiments on the popular Zhihu\footnote{https://github.com/THUIR/ZhihuRec-Dataset.} dataset, where the collection number of articles strongly reflects the item quality. Figure~\ref{fig:quality_zhihu} clearly presents the positive correlation between collection number and the post-click conversion ratio. In this figure, the value of Spearman's correlation coefficient~\cite{abdullah1990robust,hauke2011comparison} is $0.8361$ and the $p$-value is $4.4e^{-6}$, which also prove the rationality of approximating quality by the post-click conversion ratio. 

\noindent\textbf{$\bullet$ Exposure Fairness.} The exposure unfairness is that for items with same quality, more popular ones usually get more exposure.
To evaluate this unfairness, we need to quantify the item's exposure probability.
To calculate the exposure for each item in the ranking list, we refer to the position bias defined in the discounted cumulative gain~\cite{jarvelin2002cumulated}. Formally, the exposure of item $i$ can be obtain by:
\begin{equation}\label{equ:exposure}\footnotesize
\left \{
\begin{aligned}
    & v_{j} = \frac{1}{log_{2}(1 + j)}, \\
    & exp(i|P_{i}) = \sum_{j = 1}^{N} P_{i, j} * v_{j},
\end{aligned}
\right .
\end{equation}
where $v_j$ denotes the exposure probability of the $j$-th item in the ranking list, and $P_{i,j}$ is the probability of item $i$ recommended at the position of $j$. 
To evaluate the exposure fairness, we argue that the items with similar quality should receive similar exposure. As such, we divide items into groups based on item quality, and then split each group into subgroups according to different popularity. Thus, we can measure the exposure difference between these subgroups to evaluate the exposure fairness. Formally, we estimate the exposure of a subgroup $G$ by averaging the item-level exposure:
\begin{equation}\label{equ:exposure_g}\footnotesize
    Exp(G) = \frac{1}{|G|}\sum_{i \in G} exp(i|P_{i}).
\end{equation}
Thereafter, given two subgroups $G_1$ and $G_2$ with similar quality but different clicks (\ie different popularity), the exposure fairness aims to minimize the exposure disparity $D_r$ between them:
\begin{equation}\label{eq:disparity1}\footnotesize
    D_{r}(G_1, G_2) = |Exp(G_1) - Exp(G_2)|.
\end{equation}
\noindent\textbf{$\bullet$ Quality Fairness} The quality unfairness is that for items with the lower quality, a higher popularity may lead to more exposure. Following previous work~\cite{morik2020controlling}, we introduce a fairness criteria that relates exposure to item quality. 
Specifically, for the evaluation, we first estimate the average quality of groups, and then use the exposure/quality disparity between different groups to evaluate the quality unfairness. Formally,
\begin{equation}\label{eq:disparity2}\footnotesize
\left \{
\begin{aligned}
&Ratio(G) = \frac{1}{|G|} \sum_{i \in G} r_{i}, \\
&D_{l}(G_1, G_2) = \left|\frac{Exp(G_1)}{Ratio(G_1)} - \frac{Exp(G_2)}{Ratio(G_2)}\right|,
\end{aligned}
\right .
\end{equation}
where $r_i$ is the post-click conversion ratio of item $i$, and $D_{l}(G_1, G_2)$ describes the unfairness between the groups $G_1$ and $G_2$. A small disparity $D_{l}$ implies that items receive the exposure proportional to their quality.



\vspace{-5pt}
\subsection{Causal View of Unfairness}\label{ssec:causal graph}
The causal graph is a directed acyclic graph $G = \{V, E\}$, where $V$ denotes a set of variables, and $E$ represents the causal relations among variables~\cite{pearl2009causality}. 
To analyze the reasons for the two kinds of unfairness, we abstract a causal graph in Figure \ref{fig:rec_graph}(a) that describes the causal relations between item popularity, quality, and multiple user behaviors. We explain the causal graph as follows:
\begin{itemize}[leftmargin=*]
    \item $U$ and $I$ denote users and items, respectively. We assume that users can only interact with the items exposed to them, and thus $I$ represents the exposed items. 
    \item $C$ is the click probability of the user clicking the exposed item. 
    \item $L$ is the post-click probability, indicating whether the user has a post-click behavior (\eg favourite and like) on the exposed item.
    \item $Z$ represents the item popularity, which is estimated by item's received interactions.
    \item $Q$ represents the item quality, which is a hidden variable and cannot be easily quantified. As discussed in Section~\ref{sec:fairness_formulation}, we propose to approximately estimate it by the conversion ratio from clicks to post-click behaviors. If the clicked item has better quality, users should have a higher probability of liking it.
    \item $T$ represents some unobserved variables, such as the reputation and scale of item producers.
    \item $\{U, I\}\rightarrow C$: the click probability is affected by the user and exposed items. 
    \item $Z\rightarrow I$: item popularity affects the exposure of items. The recommendation models are usually trained on long-tailed data. Therefore, they tend to give popular items more exposure~\cite{germano2019few}. 
    \item $\{U, I\}\rightarrow L$: the user and exposed items determine the probability of post-click behaviors.
    \item $C\rightarrow L$: only clicked items can have post-click behaviors.
    \item $\{Z, Q\}\rightarrow L$: item popularity and quality affect the probability of post-click behaviors. Meanwhile popularity influences users' post-behaviors because of the users' herd mentality (\aka conformity)~\cite{zhang2021causal}.
    \item $Z \leftarrow T \rightarrow Q$: T affects both $Z$ and $Q$, causing the confounding effect between the two variables. For example, item producer reputation can affect the popularity of items because users tend to consume items with famous brands or well-reputed producers, making such items more likely to become popular.
    Besides, the quality of items is often affected by item producer reputation or the business scale of the producer.
    
\end{itemize}

\noindent\textbf{$\bullet$ Reason for two unfairness issues.} With the help of the causal graph in Figure \ref{fig:rec_graph}(a), we can explore the true reasons for the two kinds of unfairness. Firstly, we find that item popularity $Z$ is a confounder between $I$ and $L$, causing the first backdoor path $I\leftarrow Z\rightarrow L$. The backdoor path results in that for the same quality items, more popular ones are more likely to be exposed to the user via $Z\rightarrow I$ and receive higher post-click probabilities by $Z\rightarrow L$ due to users' herd mentality, which reflects the exposure unfairness~\cite{zhang2021causal}. 
Moreover, $T$ and $Z$ are confounders between $Q$ and $L$, causing the second backdoor path $Q\leftarrow T\rightarrow Z \rightarrow L$. Thus, item quality is correlated with popularity because of unobserved confounder $T$ via $Q\leftarrow T\rightarrow Z$. Then some low-quality items might correlate with high popularity, which will bring more exposure to such popular items. This explains the quality unfairness.

To summarize, two backdoor paths cause two kinds of unfairness in multi-behavior recommendation. To alleviate the unfairness, we need to cut off these two backdoor paths for removing the bad impact of such confounders.

\subsection{MBD Framework}\label{sec:MBD_framework}

In this work, we propose a MBD framework that leverages \textit{backdoor adjustment}~\cite{pearl2009causality} to cut off the backdoor paths and remove the bad effect caused by the confounders. Specifically, as shown in Figure \ref{fig:rec_graph}(b), we perform the intervention $do(I=i)$ by controlling the observed variable $Z$ to block the backdoor path $I \leftarrow Z \rightarrow L$. 
However, for the second backdoor path $Q \leftarrow T \rightarrow Z \rightarrow L$, $Q$ and $T$ are unobserved variables that make the backdoor adjustment infeasible. 
Thanks to the post-click conversion ratio that reflects the item quality (see Section \ref{sec:fairness_formulation}), we propose to use it for the approximation of item quality $Q$. 
Instead of controlling the unobserved $T$, we can achieve the intervention on $Q$ by controlling $Z$, which also blocks the second backdoor path. 
Besides, considering the critical impact of item quality in the recommendation, we estimate the post-click probability via $P(L|U, do(I), do(Q))$.

We denote the causal graph shown in Figure \ref{fig:rec_graph}(a) as $G$ and the intervened causal graph shown in Figure \ref{fig:rec_graph}(b) as $G'$, and then we will derive the following expression on $G$:
\begin{subequations}
\label{equ:P_do_IQ}\footnotesize
\begin{align}
&P(L|U,do(I),do(Q))=P_{G^\prime}(L| U,I,Q) \notag \\
                &=\sum_{Z}\sum_{C}{P_{G^\prime}(L| U,I,Q,Z,C)P_{G^\prime}(C|U,I,Q)}P_{G^\prime}(Z| U,I,Q) \\
                &=\sum_{Z}\sum_{C}{P_{G^\prime}(L|U,I,Q,Z,C)P_{G^\prime}(C|U,I)}P_{G^\prime}(Z) \\
                &=\sum_{Z}P_{G^\prime}(L|U,I,Q,Z,C=1)P_{G^\prime}(C=1|U,I)P_{G^\prime}(Z)+\\
                &\sum_{Z}P_{G^\prime}(L|U,I,Q,Z,C=0)P_{G^\prime}(C=0|U,I)P_{G^\prime}(Z) \\
                &\approx\sum_{Z}P_{G^\prime}(L|U,I,Q,Z,C=1)P_{G^\prime}(C=1| U,I)P_{G^\prime}(Z)\\
                &=\sum_{Z}P(L|U,I,Q,Z,C=1)P(C=1| U,I)P(Z),
\end{align}
\end{subequations}
where $P_{G'}(\cdot)$ denotes the probability function estimated by $G'$. The derivation of Equation \ref{equ:P_do_IQ} is as follows: 
\begin{itemize}[leftmargin=*]
    \item (\ref{equ:P_do_IQ}a) and (\ref{equ:P_do_IQ}c) are due to \textit{Bayes' theorem};
    \item (\ref{equ:P_do_IQ}b) is because of the $do$-calculus on $I$ and $Q$. The first backdoor path $I\leftarrow Z\rightarrow L$ is blocked by $do(I=i)$, which mitigates the exposure unfairness. The second backdoor path $Q\leftarrow T\rightarrow Z \rightarrow L$ is blocked by $do(Q=r_i)$, which mitigates the quality unfairness. Thus, $P_{G^\prime}(Z|U,I,Q)$ can be deduced as $P_{G^\prime}(Z)$. $P_{G^\prime}(C|U,I,Q)$ changes into $P_{G^\prime}(C|U,I)$ because $C$ is independent from $Q$. The user cannot know the item quality $Q$ before clicking the item;
    \item (\ref{equ:P_do_IQ}d) is because we do not consider the non-clicked data;
    \item (\ref{equ:P_do_IQ}e) is because of the same prior of $U, I, Z, Q$ on two graphs. $\{U, I, Z, Q\}\rightarrow L$ will not be changed by cutting off two paths $Z\rightarrow I$ and $T\rightarrow Q$.
\end{itemize}

Next, we need to consider how to estimate $P(L|U,do(I),do(Q))$ from the data. From the derived expression (\ref{equ:P_do_IQ}e), we need to firstly estimate $P(L|U,I,Q,Z,C=1)$ and $P(C=1|U,I)$, respectively, and then estimate $\sum_{Z}P(L|U,I,Q,Z,C=1)P(C=1|U,I)P(Z)$ to get the prediction score. The estimations are as follows:

\begin{figure}[t]
    \centering
    \includegraphics[width=.3\textwidth]{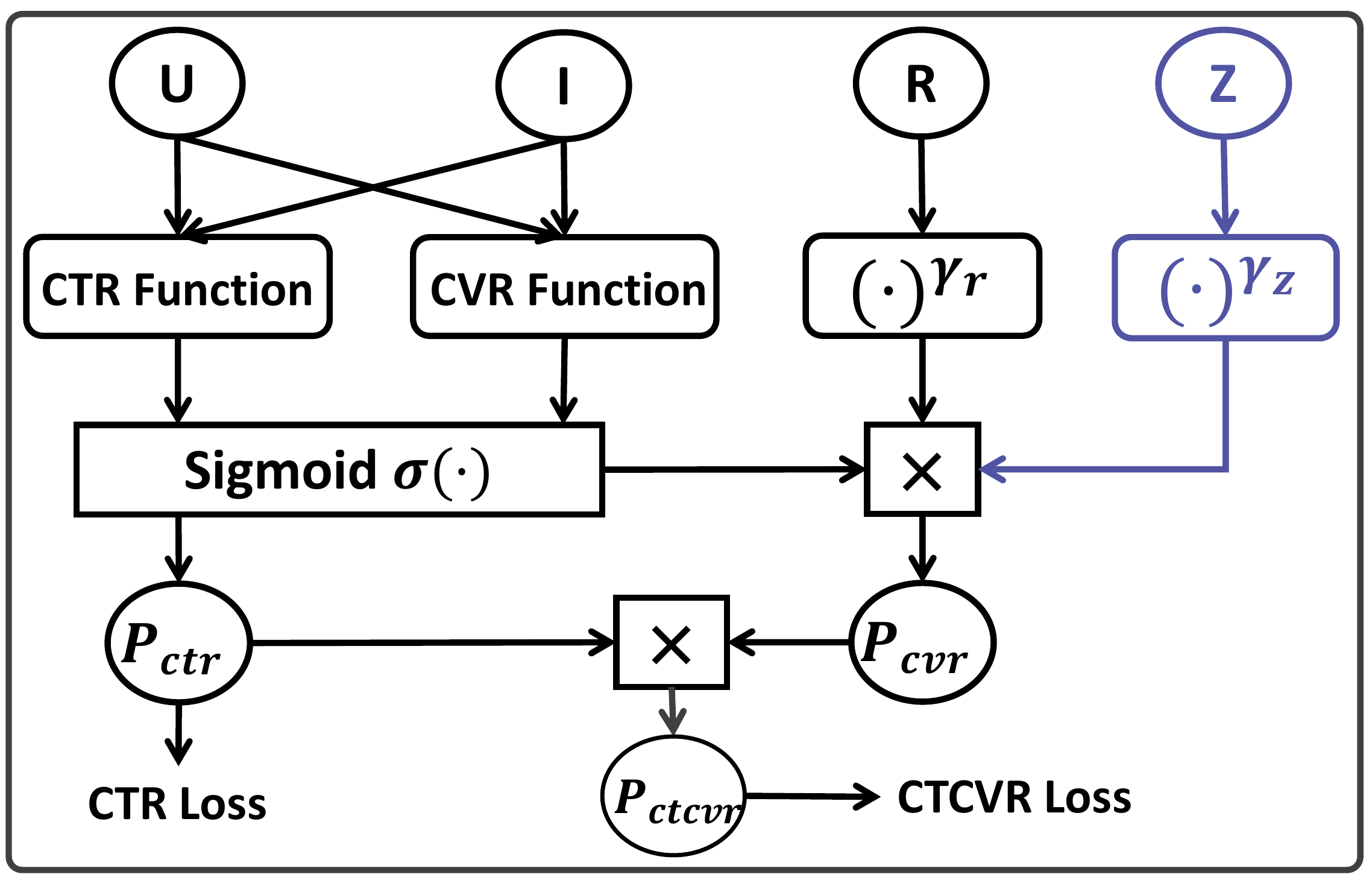} 
    \caption{The implementation of MBD. The CTR function and CVR function are the user-item matching functions $f_{c}$ and $f_{l}$, respectively. The purple arrows indicate that the item popularity $Z$ is used in the training stage but removed from the inference stage.}
    \label{fig:framework}
    \vspace{-10pt}
\end{figure}
\vspace{+5pt}
\noindent \textbf{1) Estimating $P(C=1| U,I)$.} 
This is the conditional click probability between users and items (\ie the click-through ratio (CTR)~\cite{richardson2007predicting}).
To parameterize $P(c=1|u,i)$, we use $f_c$ to represent the user-item matching function for clicks, which can be realized by any user-item matching model (\eg Matrix Factorization (MF)~\cite{koren2009matrix}). 
Hence, the probability can be calculated as:

\begin{equation}\label{equ:f_ctr}
    p_{ctr} = P(c=1|u,i) = \sigma(f_{c}(u, i)),
\end{equation}
where $\sigma$ is the sigmoid function to normalize the matching score between 0 and 1. 
We instantiate the CTR loss by the cross-entropy loss on the historical click data $\mathcal{D}$, which
consists of users' sampled unclicked items as negative samples:

\begin{equation} \label{equ:loss_ctr}
    L_{ctr}  = \sum_{(u, i)\in \mathcal{D}} - \hat{y}_clog(p_{ctr}) \\- (1-\hat{y}_c)log(1-p_{ctr})),
\end{equation}
where $\hat{y}_c\in\{0,1\}$ is the user click feedback on items. In addition, we apply $L_2$ regularization but omit it for brevity.

\vspace{+5pt}
\noindent \textbf{2) Estimating $P(L|U,I,Q,Z,C=1)$.} This conditional probability estimates that given $U=u$, $I=i$, item quality $Q=r_i$ and item popularity $Z=z_{i}$, how likely the user will have post-click behavior on the clicked item (\ie the post-click conversion ratio (CVR)). Same with CTR,
we calculate this conditional probability by:

\begin{equation}
    p_{cvr} = P(l=1|u,i,r_{i},z_{i},c=1)=\sigma(f_{l}(u,i))r_{i}^{\gamma_r}z_{i}^{\gamma_z},
\end{equation}
where $\sigma$ is the sigmoid function, and $f_l$ can be implemented by any user-item matching model. The hyper parameter $\gamma_z$ is used to smooth the item popularity and $\gamma_r$ is used to control the influence of the like/click ratio of item $i$, \ie $r_i = \frac{n_{i,l}}{n_{i,c}}$, where $n_{i,l}$ and $n_{i,c}$ are the number of post-click behaviors and clicks, respectively. 
We normalize the ratio $r$ and the popularity $z$ by the min-max normalization~\cite{al2006data} to make sure that they are the same scale. Next, we can estimate the click through\&conversion ratio (CTCVR) as:

\begin{equation}\label{equ:f_ctcvr}
\begin{aligned}
p_{ctcvr} &= P(l=1|u,i,r_{i},z_{i},c=1)*P(c=1|u,i)\\
            &=\sigma(f_l(u,i))r_{i}^{\gamma_r} z_{i}^{\gamma_z}\sigma(f_{c}(u,i)).  
\end{aligned}
\end{equation}
We calculate the CTCVR loss by the cross-entropy loss:
\begin{equation}\label{equ:loss_ctcvr}
     L_{ctcvr} = \sum_{(u, i)\in \mathcal{D}_l} - \hat{y_l}log(p_{ctcvr})\\ -(1-\hat{y}_l)log(1-p_{ctcvr}),
\end{equation}
where $\hat{y}_l\in\{0,1\}$ denotes whether the user has post-click behavior on the item, and $\mathcal{D}_l$ collects the historical post-click data. 
The overall loss $\mathcal{L}$ consists of two loss terms from the CTR and CTCVR predictions: 
\begin{equation}\label{equ:finalloss}
    \mathcal{L} = L_{ctr} + L_{ctcvr}.
\end{equation}

\vspace{+5pt}
\noindent \textbf{3) Estimating $\sum_{Z}P(L|U,I,Q,Z,C=1)P(C=1| U,I)P(Z)$.} At last, with the help of above two steps, we can estimate the interventional probability $P(L|U,do(I),do(Q))$ as follows:
\begin{subequations}\small
\label{equ:to be named}
\begin{align}\small
&P(L|U,do(I), do(Q)) \notag\\
                &=\sum_{Z}{P(L|U,I,Q,Z,C=1)P(C=1|U,I)P(Z)}  \\
                &=\sum_{Z}\sigma(f_l(u,i))r_{i}^{\gamma_r} z_{i}^{\gamma_z}\sigma(f_{c}(u,i))p(Z)   \\
                &=\sigma(f_{c}(u,i))\sigma(f_l(u,i))r_{i}^{\gamma_r}\sum_{Z} z_{i}^{\gamma_z}p(Z)   \\
                &=\sigma(f_{c}(u,i))\sigma(f_l(u,i))r_{i}^{\gamma_r}E(Z^{\gamma_z}),
\end{align}
\end{subequations}
where $E(Z^{\gamma_z})$ is the expectation of $Z^{\gamma_z}$ which is a constant and does not change the recommendation results. As Figure \ref{fig:framework} shows, we use Equation~\eqref{equ:finalloss} to train the matching functions $f_{c}(u,i) $ and $f_l(u,i)$, and model the influence of the popularity $z_i$ and quality $r_i$ on $L$ at the training stage. Thereafter, we remove the influence of the confounder $z_i$ to alleviate the two unfairness issues, \ie estimating the deconfounded prediction $s$ to rank item candidates:
\begin{equation}\label{equ:score_inference}
    s(u, i, r_{i}) = \sigma(f_{c}(u,i))\sigma(f_l(u,i))r_{i}^{\gamma_r}.
\end{equation}

%% file: 4_experiments.tex
\section{Experiments}
\subsection{Experimental Settings}
\subsubsection{Datasets}
We conduct experiments on two datasets:

1) \textbf{Kwai}: This dataset is released by Kuaishou User Interest Modeling Challenge\footnote{https://www.kuaishou.com/activity/uimc.}, which contains two-week data of click/like/follow interactions between users and videos. In this work, we utilize the click and like interactions where like is considered as the post-click behavior. We take $10$-core settings for clicks and $5$-core settings for likes since likes are relatively sparse.

2) \textbf{Tmall}: This is the dataset released in IJCAI16 challenge\footnote{https://tianchi.aliyun.com/dataset/dataDetail?dataId=53.}, which is collected from Tmall, the largest business-to-consumer E-Commerce website in China. It records two types of user behaviors (\ie click and buy) within the time period from 01/06/2015 to 30/11/2015. Buy is treated as the post-click behavior. Due to its large scale, we use 20-core settings for clicks and 10-core settings for the buys.

We summarize the statistics of datasets in Table \ref{tab:dataset_all}. For the experiments, we firstly sort the user-item interactions by timestamps, and then split the datasets into train, valid and test set with the ratio of 80\%, 10\% and 10\%.

\begin{table}[t]
\centering
\setlength{\abovecaptionskip}{0cm}
\setlength{\belowcaptionskip}{0cm}
\caption{The statistics of datasets.}
\label{tab:dataset_all}
\resizebox{.4\textwidth}{!}{
\begin{tabular}{lllll}
\hline
\textbf{Dataset} & \textbf{\#Users} & \textbf{\#Items} & \textbf{\#Clicks} & \textbf{\#Post-clicks} \\ \hline
\textbf{Kwai}    & 6,507           & 5,721           & 479,102          & 46,393                \\
\textbf{Tmall}   & 32,097          & 24,203          & 1,318,165         & 484,318               \\ \hline
\end{tabular}
}
\vspace{-5pt}
\end{table}
\begin{table*}[t]
\centering
\setlength{\abovecaptionskip}{0cm}
\setlength{\belowcaptionskip}{0cm}
\caption{The recommendation performance evaluated by post-click behaviors on the two datasets. H, R, and N denote Hit Ratio, Recall, and NDCG, respectively. The best results are highlighted in bold.
}
\setlength{\tabcolsep}{1.2mm}{
\resizebox{.9\textwidth}{!}{
\label{tab:baselines}
\begin{tabular}{c|cccccc|cccccc}
\hline
\multicolumn{1}{l|}{}   & \multicolumn{6}{c|}{\textbf{Kwai}}                                                                        & \multicolumn{6}{c}{\textbf{Tmall}}                                                                        \\ \hline
\textbf{Model}          & \textbf{H@50}   & \textbf{H@100}  & \textbf{R@50}   & \textbf{R@100}  & \textbf{N@50}   & \textbf{N@100}  & \textbf{H@50} & \textbf{H@100}  & \textbf{R@50}   & \textbf{R@100}  & \textbf{N@50}   & \textbf{N@100}  \\ \hline
\textbf{Random}         & 0.0166          & 0.0296          & 0.0122          & 0.0235          & 0.0042          & 0.0078          & 0.0050          & 0.0101          & 0.0019          & 0.0027          & 0.0006          & 0.0009          \\ \hline
\textbf{MF}             & 0.0427          & 0.0838          & 0.0215          & 0.0462          & 0.0085          & 0.0115          & 0.0866          & 0.1078          & 0.0431          & 0.0535          & 0.0312          & 0.0340          \\
\textbf{ESMM}           & 0.0577          & 0.0933          & 0.0280          & 0.0504          & 0.0093          & 0.0135          & 0.1176          & 0.1463          & 0.0617          & 0.0787          & 0.0428          & 0.0448          \\ \hline
\textbf{Multi-IPW}     & 0.0625          & 0.1155          & 0.0318          & 0.0642          & 0.0105          & 0.0166          & 0.1133          & 0.1487          & 0.0584          & 0.0801          & 0.0422          & 0.0475          \\
\textbf{Multi-IPW-DR} & 0.0561          & 0.1044          & 0.0281          & 0.0557          & 0.0092          & 0.0146          & 0.1079          & 0.1442          & 0.0546          & 0.0787          & 0.0389          & 0.0469          \\
\textbf{AdFair}    & 0.0569          & 0.0981          & 0.0289          & 0.0514          & 0.0087          & 0.0130          & 0.1144          & 0.1444          & 0.0606          & 0.0795          & 0.0433          & 0.0470          \\ 
\textbf{FER}    & 0.0624          & 0.1031          & 0.0313          & 0.0615          & 0.0101          & 0.0159          & 0.1144          & 0.1484          & 0.0616          & 0.0712          & 0.0427          & 0.0480          \\
\textbf{DCR}    & 0.0735          & 0.1251          & 0.0389          & 0.0644          & 0.0121          & 0.0183          & 0.1198          & 0.1495          & 0.0633          & 0.0794          & 0.0477          & 0.0480          \\
\textbf{MACR}      & 0.0496          & 0.0822          & 0.0252          & 0.0421          & 0.0082          & 0.0117          & 0.1143          & 0.1453          & 0.0578          & 0.0763          & 0.0439          & 0.0482          \\
\textbf{PDA}            & 0.0640          & 0.1168          & 0.0350          & 0.0684          & 0.0127          & 0.0178          & 0.0120          & 0.1511          & 0.0640          & 0.0828          & 0.0455          & 0.0491          \\ \hline
\textbf{ESMM-RI}        & 0.0625          & 0.1153          & 0.0303          & 0.0634          & 0.0101          & 0.0162          & 0.1196          & 0.1495          & 0.0630          & 0.0817          & 0.0450          & 0.0479          \\
\textbf{ESMM-RT}        & 0.0846          & 0.1281          & 0.0477          & 0.0721          & \textbf{0.0159} & 0.0204          & 0.1233          & 0.1532          & 0.0640 & 0.0839          & 0.0465          & 0.0508          \\ 
\textbf{MBD}            & \textbf{0.0909} & \textbf{0.1431} & \textbf{0.0533} & \textbf{0.0859} & 0.0153          & \textbf{0.0214} & \textbf{0.1312} & \textbf{0.1654} & \textbf{0.0698} & \textbf{0.0933} & \textbf{0.0484} & \textbf{0.0536} \\ \hline
\end{tabular}
}
}
\end{table*}
\vspace{-10pt}

\subsubsection{Baselines}
We compare our proposed MBD framework with the following baselines:

\textbf{- Random.} This method simply recommends the randomly sampled items for all users.

\textbf{- MF~\cite{koren2009matrix}.} This is the basic matrix factorization model.

\textbf{- ESMM~\cite{ma2018entire}} is a multi-task learning method for multi-behavior recommendation. It learns the CTR and CVR prediction models simultaneously and uses the CTR and CTCVR losses for training. In this work, we take the CTCVR score as the post-click behavior prediction score. In addition, we use MF to make the CTR prediction and Multi-layer Perceptron (MLP) to make the CVR prediction, where the hidden dimensions are 128 and 64; and the activation function is sigmoid.

\textbf{- Multi-IPW-DR~\cite{zhang2020large}} uses Multi-task Inverse Propensity Weighting (Multi-IPW) estimator and Multi-task Doubly Robust (Multi-DR) estimator to mitigate selection bias and data sparsity in multi-behavior recommendation. We tune the IPW bound percentage $\tau$ in the range of $\{0.5,0.55,0.6,...,0.95\}$. \textbf{Multi-IPW} is the method that only uses Multi-task IPW estimator.

\textbf{- AdFair~\cite{li2021towards}} uses an adversarial training framework to generate feature-independent user embeddings for recommendation and achieve personalized fairness for users. We use ESMM as its user-item matching model and take the popularity as the sensitive feature of items. We tune the adversarial parameter $\lambda$ in the range of $\{0.01,0.1,...,100\}$.

\textbf{- FER~\cite{singh2018fairness}} proposes a conceptual and computational framework that applies fairness constraints on the item rankings. It uses the average relevance of each document (normalized between 0 and 1) with the query as the utility. In the experiments, we use the user-item relevance as the utility of the item.

\textbf{- DCR~\cite{he2023addressing}} proposes a Deconfounding Causal Recommendation (DCR) framework to estimate the causal effect of content features to remove the backdoor path behind user-item matching, with a MoE architecture to speed up the inference process.  

\textbf{- MACR~\cite{wei2021model},} which is model-agnostic using counterfactual reasoning method for eliminating popularity bias. In this study, we use ESMM as its backbone model. We tune the hyper-parameters $\lambda$ and $\beta$ in the range of $\{0.001,0.01,...,1\}$, which control the weights of prediction loss, and tune the parameter $c$ for inference in the range of $\{0.01,0.05,0.1...,1\}$.

\textbf{- PDA~\cite{zhang2021causal}, } which use causal intervention for mitigating popularity bias and use the adjusted popularity in inference to get better recommendation accuracy.

\textbf{- ESMM-RI} and \textbf{ESMM-RT} are two varieties of ESMM by considering item quality in recommendation. ESMM-RI utilizes the well-trained ESMM but multiplies the post-click conversion \textbf{R}atio with  CTCVR prediction score during the \textbf{I}nference stage. This will directly leverage the information of item quality for recommendations. Different from ESMM-RI, ESMM-RT multiples the post-click conversion \textbf{R}atio with the CTCVR prediction score in both \textbf{T}raining and inference stages.

\subsubsection{Hyper-parameters.} For a fair comparison, all methods are tuned on validation data. We optimize all models by Adam optimizer with the batch size of 4,096 and learning rate of $0.01$ for all the experiments. $L_2$ regularization coefficient is searched from the range of $\{0.001,...,10^{-6}\}$. Besides, we adopt the early stopping strategy that stops training if top-$50$ Recall plus Hit Ratio on the validation dataset does not increase for ten epochs. For the parameter $\gamma_{r}$ and $\gamma_{z}$, we tune them in the range of $\{0.1, 0.2,...,2\}$ with step 0.1. Larger $\gamma_z$ and $\gamma_r$ mean larger impact of the popularity and quality of items. 

\subsubsection{Metrics.}\label{metrics}
To measure the recommendation performance, we adopt three widely-used evaluation metrics: Hit Ratio, Recall and NDCG. All metrics are computed by the all-ranking protocol --- all items that are not interacted by a user are the candidates. We take the recommendation accuracy of post-click behavior as the evaluation results of different models. Because the post-click behaviors are so sparse in datasets and the quantity of items is numerous that the difference of models' performance on top-5 or top-10 is not significant. Therefore, we report the recommendation results of $K=50$ and $K=100$. 

\begin{table*}[t]
\centering
\setlength{\abovecaptionskip}{0cm}
\setlength{\belowcaptionskip}{0cm}
\caption{The disparity on the exposure unfairness by different methods on two datasets. The best results except Random are highlighted in bold. Smaller scores indicate better performance.}
\label{tab:fairness1}
\begin{center}
\setlength{\tabcolsep}{0.8mm}{
\resizebox{.9\textwidth}{!}{%
\begin{tabular}{c|ccccc|ccccc}
\hline
\multicolumn{1}{l|}{} & \multicolumn{5}{c|}{\textbf{Kwai}}                                                                                                                                                                      & \multicolumn{5}{c}{\textbf{Tmall}}                                                                                                                                                                    \\ \hline
\textbf{Model}                                & \multicolumn{1}{l}{\textbf{{[}0.1, 0.14)}} & \multicolumn{1}{l}{\textbf{{[}0.14,0.18)}} & \multicolumn{1}{l}{\textbf{{[}0.18,0.22)}} & \multicolumn{1}{l}{\textbf{{[}0.22,0.26)}} & \multicolumn{1}{l|}{\textbf{{[}0.26,0.30)}} & \multicolumn{1}{l}{\textbf{{[}0.6,0.64)}} & \multicolumn{1}{l}{\textbf{{[}0.64,0.68)}} & \multicolumn{1}{l}{\textbf{{[}0.68,0.72)}} & \multicolumn{1}{l}{\textbf{{[}0.72,0.76)}} & \multicolumn{1}{l}{\textbf{{[}0.76,0.80)}} \\ \hline
\textbf{Random}                               & 0.0013                                     & 0.0014                                     & 0.0029                                     & 0.0029                                     & 0.0066                                      & 0.0016                                    & 0.0021                                     & 0.0017                                     & 0.0024                                     & 0.0044                                     \\ \hline
\textbf{ESMM}                                 & 0.0185                                     & 0.0163                                     & 0.0248                                     & 0.0266                                     & 0.0335                                      & 0.0205                                    & 0.0221                                     & 0.0210                                     & 0.0239                                     & 0.0175                                     \\
\textbf{Multi-IPW}                            & 0.0125                                     & 0.0101                                     & 0.0419                                     & 0.0407                                     & 0.0245                                      & 0.0110                                    & 0.0113                                     & 0.0323                                     & 0.0176                                     & 0.0178                                     \\
\textbf{AdFair}                               & 0.0091                                     & 0.0193                                     & 0.0205                                     & 0.0213                                     & 0.0234                                      & 0.0102                                    & 0.0188                                     & 0.0130                                     & 0.0298                                     & 0.0321                                     \\
\textbf{FER}                                  & 0.0131                                     & 0.0090                                     & 0.0450                                     & 0.0290                                     & 0.0213                                      & 0.0116                                    & 0.0131                                     & 0.0252                                     & 0.0177                                     & 0.0177                                     \\
\textbf{MACR}                                 & 0.0171                                     & 0.0194                                     & 0.0392                                     & 0.0194                                     & 0.0310                                      & 0.0124                                    & 0.0228                                     & 0.0198                                     & 0.0274                                     & 0.0183                                     \\
\textbf{PDA}                                  & 0.0020                                     & 0.0094                                     & 0.0135                                     & 0.0184                                     & 0.0093                                      & 0.0099                                    & 0.0173                                     & 0.0146                                     & 0.0166                                     & 0.0193                                     \\
\textbf{ESMM-RT}                              & 0.0021                                     & 0.0053                                     & 0.0211                                     & 0.0193                                     & 0.0162                                      & 0.0065                                    & 0.0072                                     & 0.0127                                     & \textbf{0.0122}                            & 0.0125                                     \\ \hline
\textbf{MBD}                                  & \textbf{0.0018}                            & \textbf{0.0046}                            & \textbf{0.0109}                            & \textbf{0.0171}                            & \textbf{0.0084}                             & \textbf{0.0062}                           & \textbf{0.0043}                            & \textbf{0.0107}                            & 0.0144                                     & \textbf{0.0121}                            \\ \hline
\end{tabular}
}
}
\end{center}
\end{table*}

\subsection{Performance on Accuracy}
In this section, we will study the recommendation performance on accuracy of our proposed method compared to other baselines. Table \ref{tab:baselines} shows top-$K$ recommendation results evaluated by reliable post-click behaviors on two datasets, and the main observations are as follows:
\begin{itemize}[leftmargin=*]
    \item The proposed MBD achieves the best performance and outperforms all the baselines on the two datasets.
    Compared to ESMM-RT, the H$@50$ improvements on two datasets are $7.45\%$ and $6.41\%$, and the R$@50$ improvements are $11.7\%$ and $9.06\%$, respectively.
    Our MBD framework improves the recommendation accuracy through mitigating the two unfairness. By alleviating the exposure unfairness, we remove the interference of popularity which hinders the model to learn the true interests of users. 
    Furthermore, by alleviating the quality unfairness, the interference of low-quality but high popularity items on the user's choice is removed. In the end, we can better fit the user's true interests and improve the recommendation accuracy.
    That verifies the effectiveness of MBD because it takes into account the item quality and removes the bad effect of popularity bias through two interventions.
    
    \item 
    MF only performs better than the Random method. 
    That is because MF only uses the post-click historical data for model learning and suffers from the data sparsity issue~\cite{ma2018entire}, which hinders the model to learn user interests well.
    
    \item ESMM performs better than MF, MACR, and AdFair. It uses multi-task learning strategy to learn both CTR and CVR tasks simultaneously, which effectively solves the data sparsity issue and enables the model to recommend more satisfying items.
    
    \item Multi-IPW-DR performs worse than Multi-IPW because the imputation model aims to estimate the error for the missing data in the whole exposure space. 
    However, in this work, we assume that the unclicked data in the exposure space are not available for training. 
    Therefore, the imputation model does not work well. In addition, the performance of Multi-IPW is unstable. It performs better than ESMM on Kwai but worse on Tmall. This may be attributed to the high model variance of the propensity score, which decreases the robustness of IPW methods. 
    
    \item AdFair performs worse than ESMM, which is because AdFair makes a trade-off between recommendation accuracy and fairness through adversarial learning. Besides, the model is easy to focus on fairness at the expense of large accuracy because it blindly masks the items' features without considering the causal relationships between different factors in the recommendation.
    
    \item FER performs worse than ESMM-RI, ESMM-RT and PDA on both datasets because the model constrains the click-through rate for the two groups to be proportional to their average utility without considering the quality of items and the relations of multiple behaviors.

    \item DCR has better performance than ESMM-RI because it incorporates MoE (Mixture of Experts) method in its model design. The overall results still show the better performance of our proposed method than DCR. It might because that the DCR model only focus on click-through rate in recommendation and ignores the quality of item and the relation of multiple behaviors.
    
    \item MACR has inferior performance than ESMM on two datasets. The reason is that it removes the natural direct effect~\cite{pearl2009causality} of items and users for mitigating the influence of item popularity and user conformity without considering other important factors like item quality. This may hinder the model from learning accurate user interests from biased interactions.
    
    \item PDA shows better results than MACR, Multi-IPW, and AdFair. By doing intervention on items and adjusting popularity bias during the inference stage, PDA gets better recommendation accuracy. Nevertheless, it still performs worse than ESMM-RT and MBD because of ignoring the relations between item quality and popularity.
    \item Our proposed variants ESMM-RT and ESMM-RI surpass ESMM, MACR, and AdFair because these two methods take into account the item quality. ESMM-RI performs worse than ESMM-RT because it does not utilize the post-click conversion ratio for training and only multiplies it with the prediction score for inference, which may hurt some well-learned user interests during training.
\end{itemize}

\subsection{Performance on Fairness}\label{sec:fair}
In this section, we conduct an analysis of recommendation results for each kind of unfairness to verify the effectiveness of our proposed method on mitigating the two unfairness issues.



\subsubsection{Analysis on the exposure unfairness.} 
\begin{figure}[t]
    \center
    \includegraphics[width=.4\textwidth]{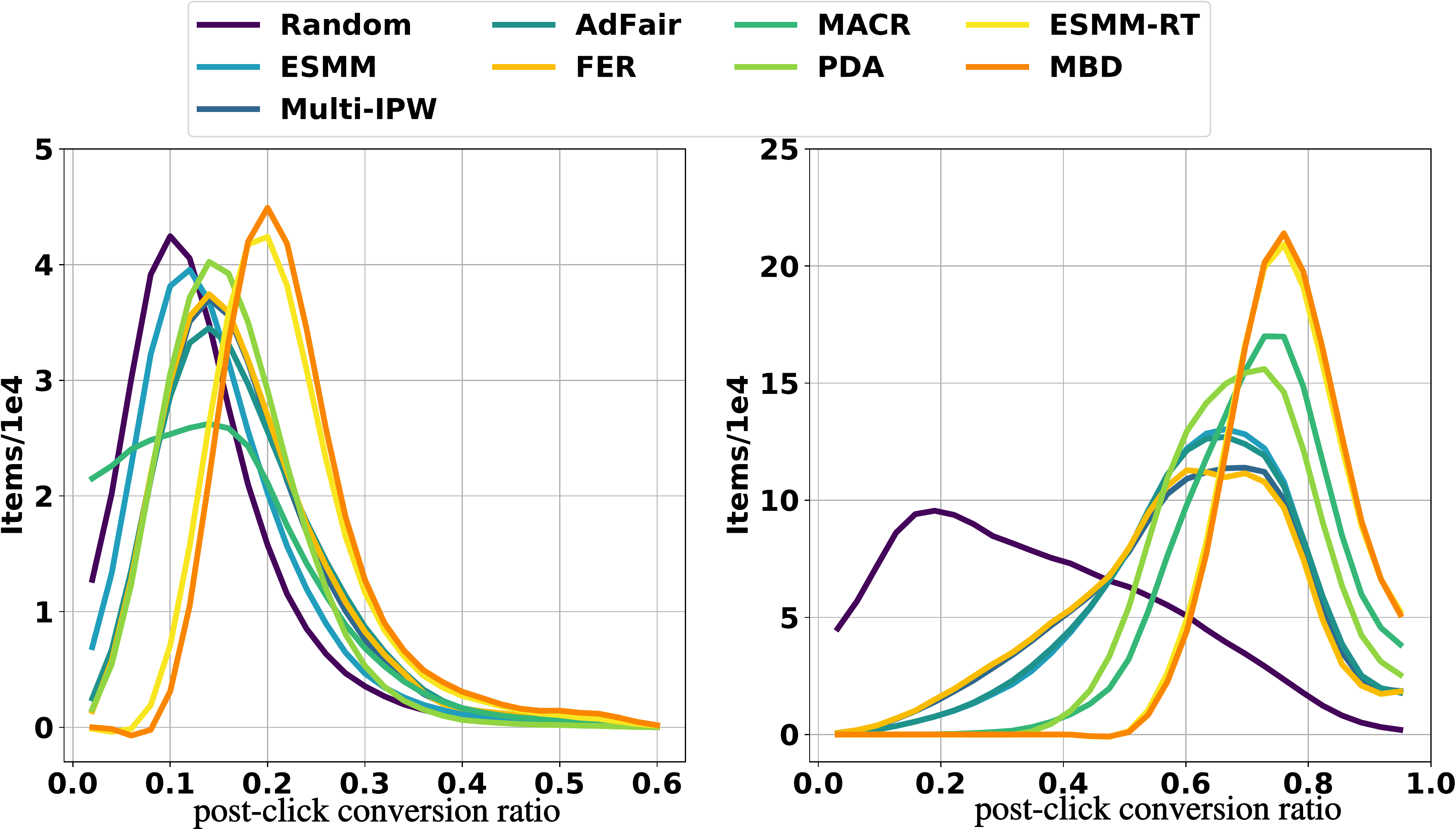} 
    \caption{Distribution of top-$50$ recommended items \wrt ratio on the two datasets. The first sub-figure shows the distribution on Kwai and the second one shows that on Tmall.}
    \label{Figure:distribution-ratio}
\end{figure}

In this section, we divide all users' top-$50$ recommendations into several groups by two steps: 1) we divide the items into several groups by their post-click conversion ratios. The items in the same ratio range are likely to have a similar quality. As shown in Figure \ref{Figure:distribution-ratio}, we choose the ratio threshold by ensuring that the quantity of recommended items in each group is similar. 2) We further separate each group into subgroups by items' received clicks. Similarly, each subgroup has the same quantity of recommendation frequency to reduce the massive difference of items' distribution between the subgroups.

\begin{figure}[t]
\centering
\subfigure[Item distribution \wrt likes on Kwai.]
{\label{Figure:a}\includegraphics[width=0.2\textwidth]{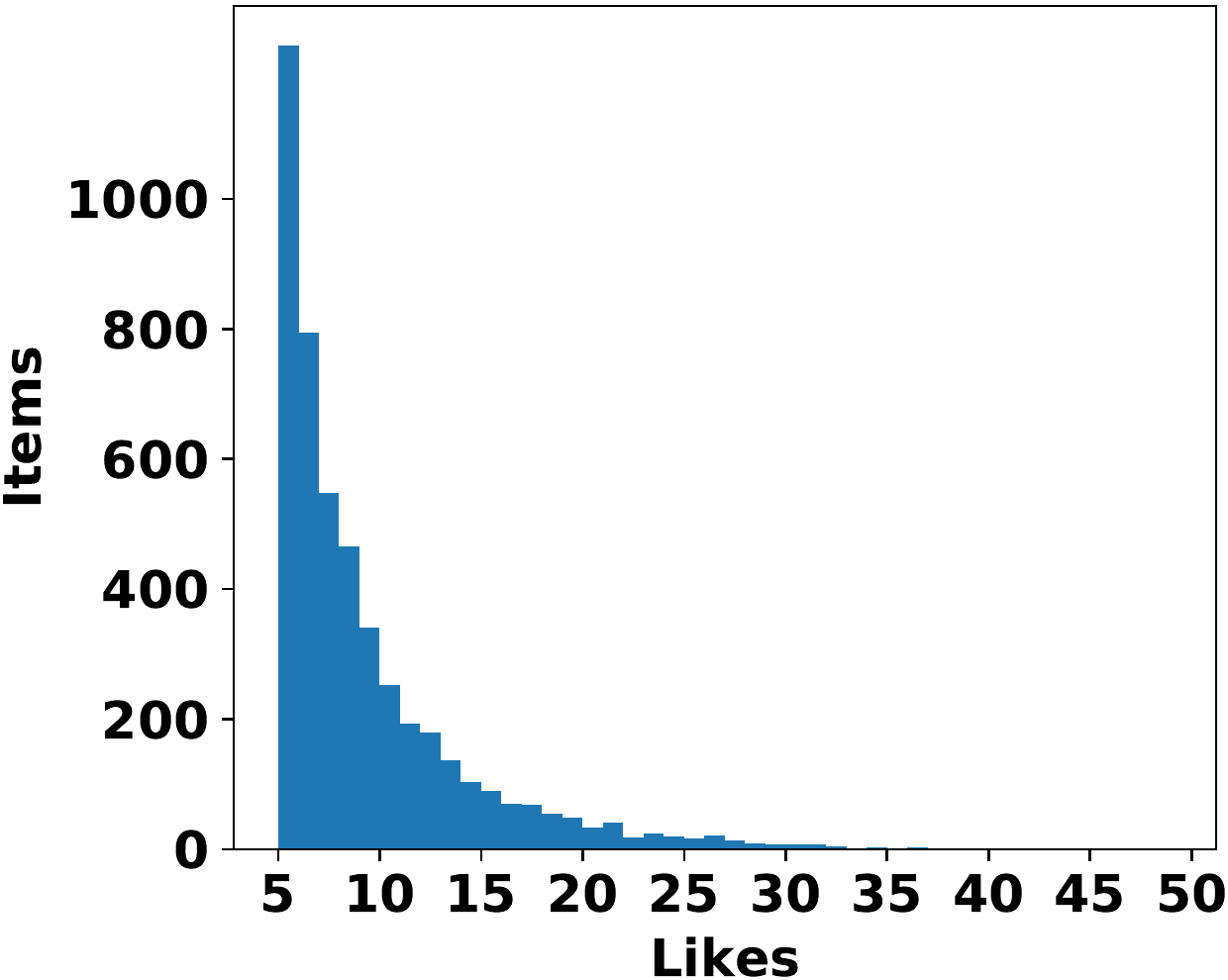}}
\subfigure[Item distribution \wrt buys on Tmall.]
{\label{Figure:b}\includegraphics[width=0.2\textwidth]{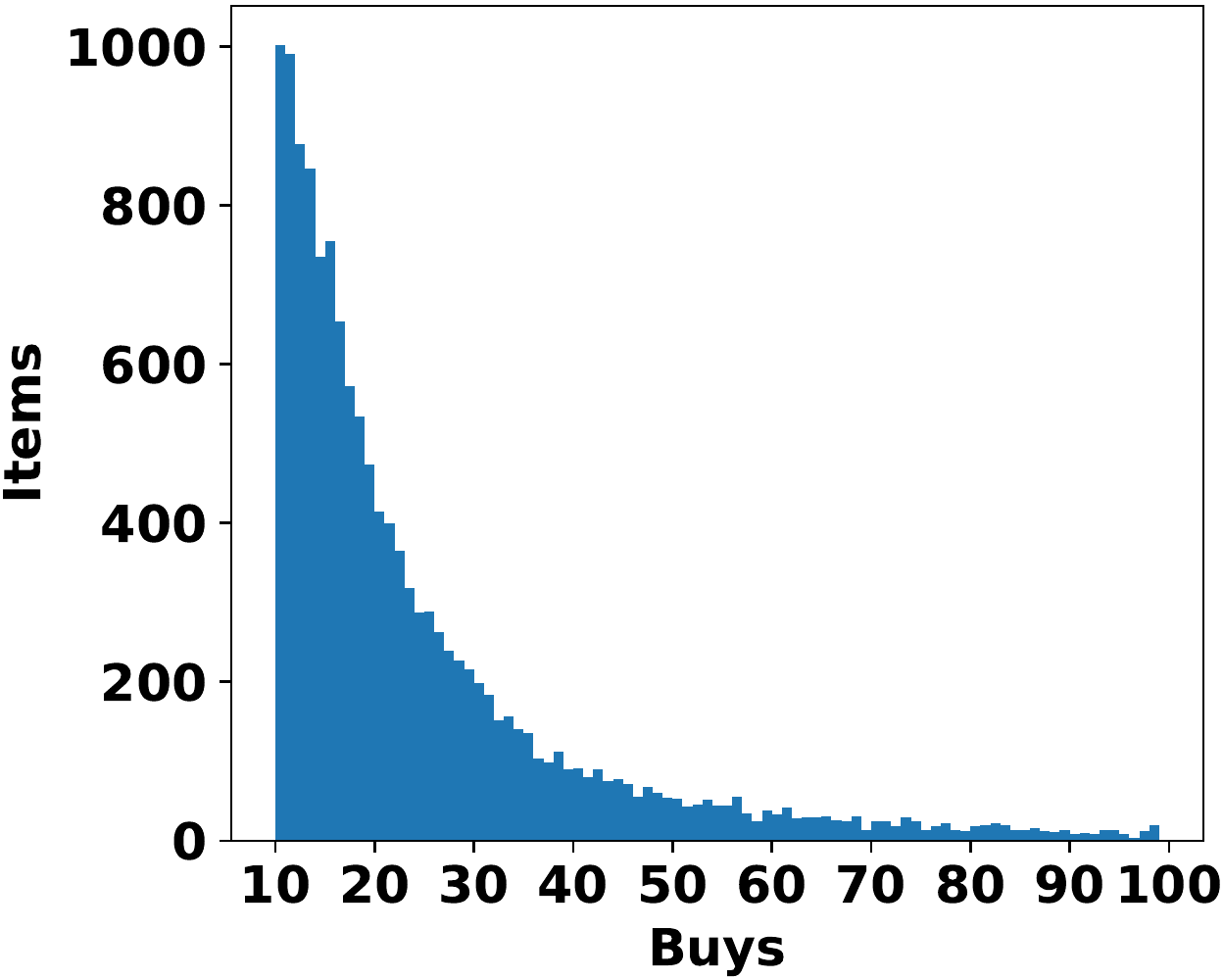}}
\caption{Distribution of all items \wrt post-click behaviors on the two datasets.}
\vspace{-10pt}
\label{Figure:distribution-postclick}
\end{figure}

\begin{figure}[t]
\vspace{-10pt}
    \centering
    \includegraphics[width=.45\textwidth]{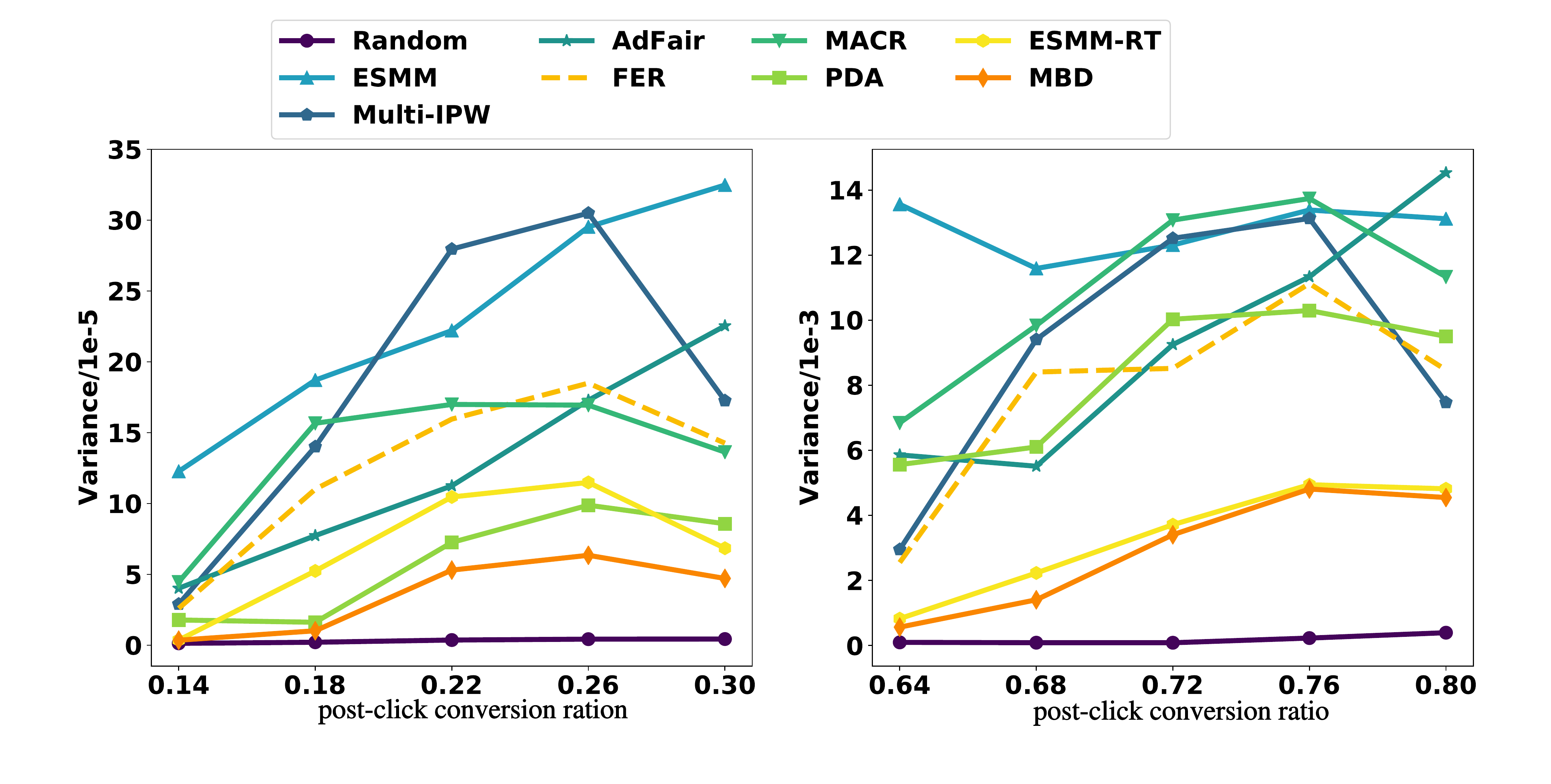} 
    \caption{Variance of exposure in each group on two datasets. The first sub-figure shows the variance on Kwai and the second one shows the variance on Tmall.}
    \label{Figure:fairness1}
\end{figure}

\begin{figure}[h]
    \centering
    \includegraphics[width=.45\textwidth]{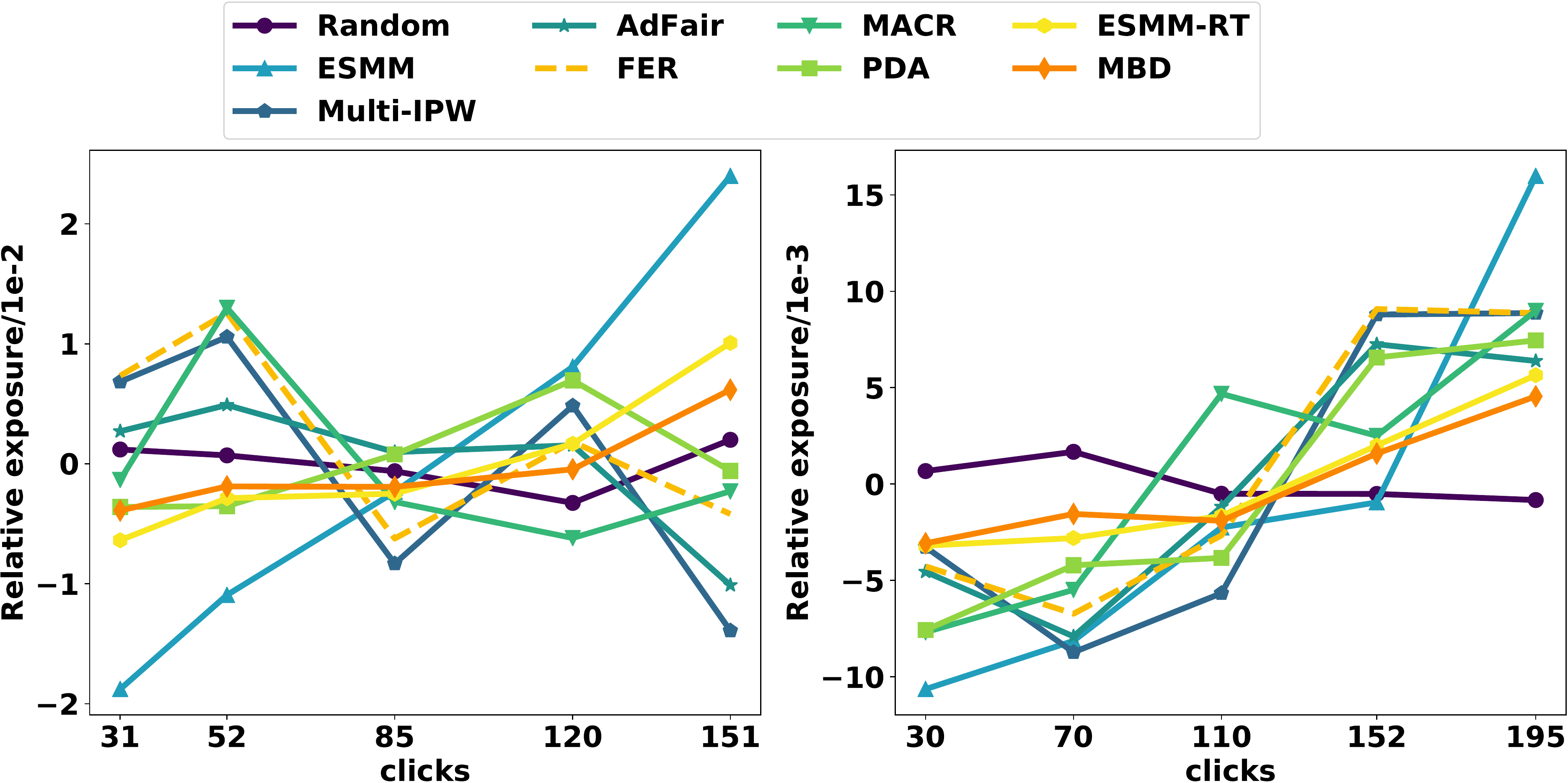} 
    \caption{Subgroups' exposure \wrt clicks on the two datasets. The first sub-figure shows the results on Kwai and the second one is on Tmall. We show the group with the most items whose ratio is in $[0.18, 0.22)$ for Kwai and $[0.68,0.72)$ for Tmall. To better compare the trends of different curves, we visualize the relative exposure by subtracting an average value in each item group.}
    \label{Figure:fairness1-detail}
    \vspace{-10pt}
\end{figure} 

We calculate the absolute difference of exposure between every two subgroups by Equation~\eqref{eq:disparity1} and take the average as the disparity of that group. The disparities of all groups are shown in Table \ref{tab:fairness1}. If the disparity is closer to zero, the exposure of items is less relevant to the item's popularity, which denotes fairer recommendations for the group. 
Besides, in Figure \ref{Figure:fairness1}, we examine the subgroup variance of the exposure in each group, which intuitively shows the exposure difference between the subgroups. 

From Table \ref{tab:fairness1} and Figure \ref{Figure:fairness1}, we find that the Random method achieves the lowest variance because it is not affected by item popularity in the historical data. Furthermore, MBD framework gets better performance as compared to other methods. It validates that MBD mitigates the disparity in the subgroups with different popularity and treats them equally. The IPW method exhibits high exposure variance in some groups due to its unstable propensity estimation. 

To examine the recommendation results with more details, we visualize the exposure of each subgroup in the group with the largest item number, which is shown in Figure \ref{Figure:fairness1-detail}:
\begin{itemize}[leftmargin=*]
    \item The curve of the random method is the most horizontal one, which means that the item popularity has no effect on the recommendation results. 
    \item The curve of MBD is the smoothest one except the random method, which implies that the exposure of items hardly changes with popularity. It verifies the effectiveness of MBD in mitigating the exposure unfairness. 
    \item For ESMM, the exposure of subgroups significantly increases with clicks, which reflects that the items with high popularity are receiving more exposure. This is consistent with the analysis of the exposure unfairness.
    \item Multi-IPW, AdFair, MACR, and PDA are all helpful to mitigate the influence of popularity bias. However, they still can be improved by explicitly considering the item quality for fairness.
    \item In Figure \ref{Figure:fairness1-detail}, the fairness improvements of MBD on Kwai are better than those on Tmall because the user behaviors on the E-commerce platform are less affected by popularity bias. As shown in Figure \ref{Figure:distribution-ratio}, the post-click conversion ratio is larger on Tmall than Kwai, which means that the user behaviors are more demand-oriented on Tmall, \ie users are more likely to click what they want to buy on the e-commerce platform. Therefore, the users' interactions on Tmall are less affected by popularity bias than Kwai, and thus these debiasing methods have smaller advantages on Kwai.
\end{itemize}

\begin{table*}[]
\centering
\setlength{\abovecaptionskip}{0cm}
\setlength{\belowcaptionskip}{0cm}
\caption{The disparity on the quality unfairness on two datasets. The best results are highlighted in bold.}
\label{tab:fairness2}
\resizebox{0.9\textwidth}{!}{%
\begin{tabular}{c|ccccc|cccccc}
\hline
                   & \multicolumn{5}{c|}{\textbf{Kwai}}                              & \multicolumn{6}{c}{\textbf{Tmall}}                                                                        \\ \hline
\textbf{Model}     & \textbf{5}      & \textbf{6}      & \textbf{7}      & \textbf{8}      & \textbf{9}      & \textbf{10-15}  & \textbf{14-17}  & \textbf{18-21}  & \textbf{22-25}  & \textbf{26-29}  & 30-40                                   \\ \hline
\textbf{Random}    & 0.4525          & 0.5503          & 0.5630          & 0.5678          & 0.5809          & 0.4932          & 0.1073          & 0.1034          & 0.0844          & 0.0926          & 0.1026          \\ \hline
\textbf{ESMM}      & 0.4512          & 0.5079          & 0.6263          & 0.5250          & 0.5832          & 0.6040          & 0.0751          & 0.0890          & 0.0876          & 0.0751          & 0.0691          \\
\textbf{Multi-IPW} & 0.3860          & 0.5025          & 0.5421          & 0.3789          & 0.5524          & 0.3413          & 0.0730          & 0.0805          & 0.0793          & 0.0724          & 0.0818          \\
\textbf{AdFair}    & 0.5675          & 0.6872          & 0.6306          & 0.6759          & 0.7241          & 0.3750          & 0.0856          & 0.0755          & 0.0790          & 0.0787          & 0.0708          \\
\textbf{FER}       & 0.4387          & 0.4846          & 0.5094          & 0.3989          & 0.4414          & 0.3583          & 0.0793          & 0.0781          & 0.0758          & 0.0700          & 0.0679          \\
\textbf{MACR}      & 0.3438          & 0.3868          & 0.5613          & 0.4269          & 0.3926          & 0.2519          & 0.0578          & 0.0530          & 0.0654          & 0.0633          & 0.0607          \\
\textbf{PDA}       & 0.4840          & 0.4845          & 0.5670          & 0.5025          & 0.4184          & 0.4317          & 0.0552          & 0.0515          & 0.0556          & 0.0536          & 0.0508          \\
\textbf{ESMM-RT}   & 0.2248          & 0.2023          & 0.2442          & 0.1906          & 0.2597          & 0.1797          & 0.0197          & 0.0181          & 0.0199          & \textbf{0.0198} & 0.0343          \\ \hline
\textbf{MBD}       & \textbf{0.1600} & \textbf{0.1476} & \textbf{0.2263} & \textbf{0.1471} & \textbf{0.1848} & \textbf{0.1409} & \textbf{0.0155} & \textbf{0.0141} & \textbf{0.0179} & 0.0212          & \textbf{0.0252} \\ \hline
\end{tabular}
}
\vspace{-10pt}
\end{table*}

\begin{figure}[t]
    \centering
    \includegraphics[width=.45\textwidth]{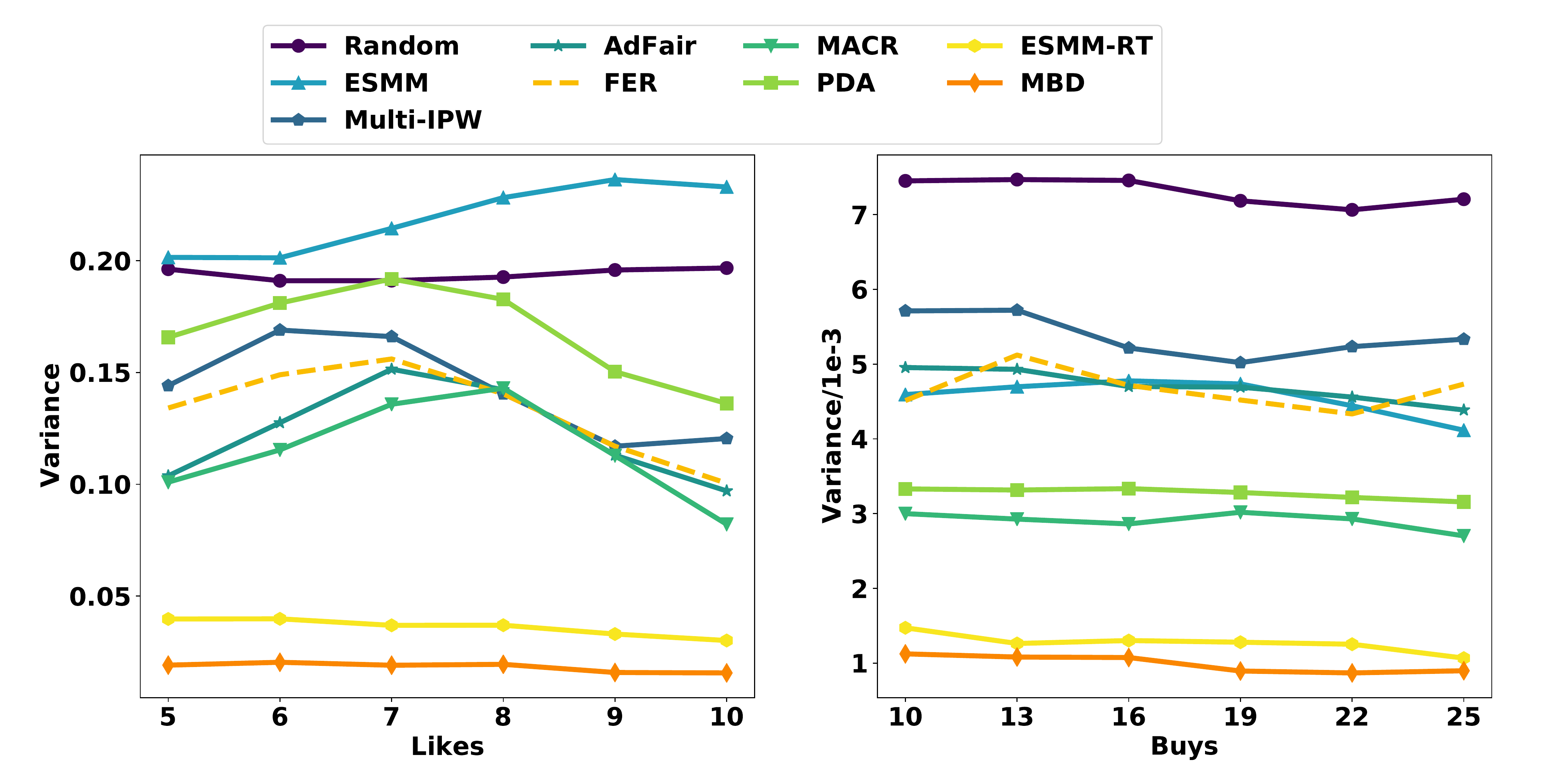} 
    \caption{Variance of exposure divided by ratio in each group on two datasets. The first sub-figure shows the variance on Kwai and the second presents the variance on Tmall.}
    \label{Figure:fairness2}
\end{figure}

\begin{figure}[ht]
    \centering
    \includegraphics[width=.45\textwidth]{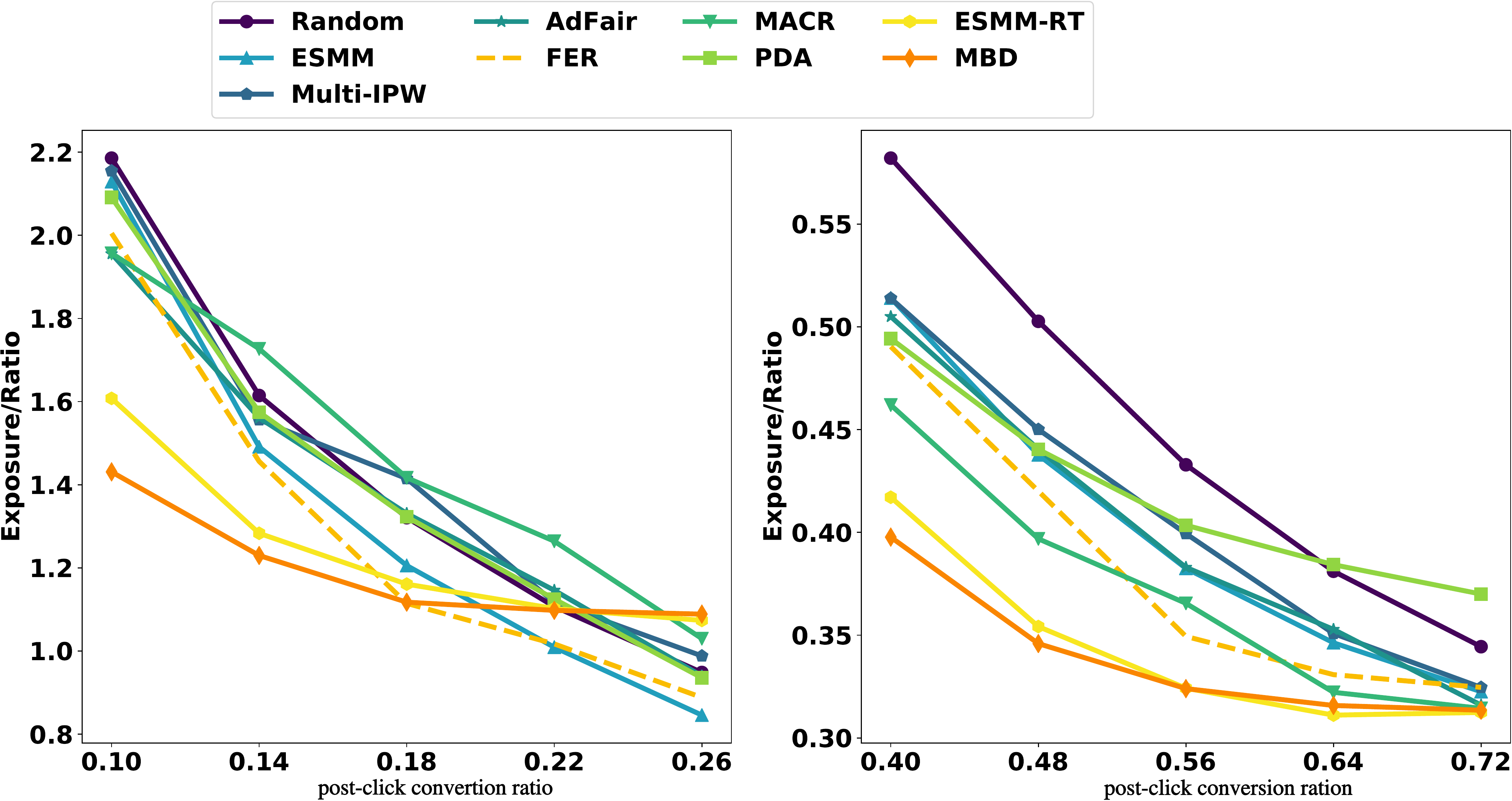} 
    \caption{Subgroup's exposure/ratio values \wrt different ratios on two datasets. The first sub-figure shows the results on Kwai and the second reports the performance on Tmall. We show the group with the most items, whose post-click number is $5$ for Kwai and $[10,13)$ for Tmall.}
    \label{Figure:fairness2-detail}
    \vspace{-10pt}
\end{figure}

\subsubsection{Analysis on the quality unfairness} We separate all users' top-$50$ recommendations into several groups by two steps: 1) we divide the items into groups by their received post-clicks. 
The thresholds are chosen to ensure that the items in each groups are relatively even. The distributions of post-clicks are shown in Figure \ref{Figure:distribution-postclick}. 
2) We separate each group into several subgroups by items' post-click conversion ratios, which indicates that the items in each subgroup have similar quality.
In each group, we calculate the disparity between every two subgroups by Equation~\eqref{eq:disparity2} and take the average as the disparity of that group. The exposure disparities regarding item quality are shown in Table \ref{tab:fairness2}. 
Besides, as in Figure~\ref{Figure:fairness2}, we examine the variance of the exposure divided by the ratio in each group.

From Table~\ref{tab:fairness2} and Figure~\ref{Figure:fairness2}, we observe that our proposed MBD framework achieves the lowest disparity and variance as compared to other methods, which illustrates that MBD will match the corresponding exposure based on the item quality.
In addition, AdFair, MACR, and PDA perform better than ESMM, indicating that they mitigate the quality unfairness to some extent by eliminating the influence of popularity. However, they do not achieve superior performance due to ignoring the influence between the item quality and popularity.

To see more details in the recommendation results, we examine the results of each subgroup in the group with the maximal item number. As shown in Figure \ref{Figure:fairness2-detail}, x-axis denotes the post-click conversion ratio of each subgroup while y-axis is the value of exposure divided by the post-click conversion ratio (illustrated in Equation \eqref{eq:disparity2}). From Figure \ref{Figure:fairness2-detail}, we can find that:
\begin{itemize}[leftmargin=*]
    \item The curve of MBD is the flattest one, which shows that the items are more likely to receive exposure according to their quality. This justifies the effectiveness of MBD in mitigating the quality unfairness.
    \item On the curve of ESMM, the exposure/ratio value over subgroups decreases with ratio, which means that for items with similar post-click behaviors, the ones with low qualities have received more exposure to such items because of the high popularity. 
    \item Multi-IPW, AdFair, MACR and PDA mitigate the influence of popularity bias because their curves are more horizontal as compared to the one of ESMM.
\end{itemize}

\subsubsection{Ablation study}
In this section, we will disable two interventions separately and verify the effectiveness of each intervention. As shown in Equation (\ref{equ:doI}), if we only conduct the intervention on $I$, the prediction probability will multiply the average popularity of the same-quality items in the inference stage. In Equation (\ref{equ:doQ}), if we only intervene on $Q$, the prediction probability will multiply the popularity of the item in inference.
\begin{subequations}
\label{equ:doI}\footnotesize
\begin{align}
&P(L|U,do(I), Q) \notag\\
                &=\sum_{Z}{P(L|U,I,Q,Z,C=1)P(C=1|U,I)P(Z|Q)}  \\
                &=\sum_{Z}\sigma(f_l(u,i))r_{i}^{\gamma_r} z_{i}^{\gamma_z}\sigma(f_{c}(u,i))p(Z|Q)   \\
                &=\sigma(f_{c}(u,i))\sigma(f_l(u,i))r_{i}^{\gamma_r}\sum_{Z} z_{i}^{\gamma_z}p(Z|Q)
\end{align}
\end{subequations}
\begin{subequations}
\label{equ:doQ}\footnotesize
\begin{align}
&P(L|U, I, do(Q))\notag\\
                &=\sum_{Z}{P(L|U,I,Q,Z,C=1)P(C=1|U,I)P(Z|I)}  \\
                &=\sum_{Z}\sigma(f_l(u,i))r_{i}^{\gamma_r} z_{i}^{\gamma_z}\sigma(f_{c}(u,i))p(Z|I)   \\
                &=\sigma(f_{c}(u,i))\sigma(f_l(u,i))r_{i}^{\gamma_r}\sum_{Z} z_{i}^{\gamma_z}p(Z|I)
\end{align}
\end{subequations}

\begin{table}[t]
\setlength{\abovecaptionskip}{0.10cm}
\caption{ Ablation study of separately disabling each intervention on Kwai.}
\label{tab:ablation}
\resizebox{.5\textwidth}{!}{
\begin{tabular}{c|cccccc}
\hline
\multicolumn{1}{l|}{} & \multicolumn{6}{c}{\textbf{Unfairness 1}}                                                                                                                         \\ \hline
\diagbox[]{\textbf{Model}}{\textbf{Ratio}}        & \textbf{{[}0.1,0.14)} & \textbf{{[}0.14,0.18)} & \textbf{{[}0.18,0.22)} & \textbf{{[}0.22,0.26)} & \multicolumn{1}{c|}{\textbf{{[}0.26,0.30)}} & \textbf{Average} \\ \hline
\textbf{ESMM-RT}      & 0.0021               & 0.0053                & 0.0211                & 0.0193                & \multicolumn{1}{c|}{0.0162}                & 0.0128           \\
\textbf{ESMM-doQ}     & 0.0024               & 0.0072                & 0.0119                & 0.0179                & \multicolumn{1}{c|}{0.0147}                & 0.0108           \\
\textbf{ESMM-doI}     & 0.0019               & 0.0042                & 0.0116                & 0.0179                & \multicolumn{1}{c|}{0.0142}                & 0.0096           \\
\textbf{MBD}          & 0.0018               & 0.0046                & 0.0109                & 0.0171                & \multicolumn{1}{c|}{0.0084}                & 0.0085           \\ \hline
\textbf{}             & \multicolumn{6}{c}{\textbf{Unfairness 2}}                                                                                                                         \\ \hline
\diagbox[]{\textbf{Model}}{\textbf{Likes}}        & \textbf{6}            & \textbf{7}             & \textbf{8}             & \textbf{9}             & \multicolumn{1}{c|}{\textbf{10-15}}                              & \textbf{Average} \\ \hline
\textbf{ESMM-RT}      & 0.2248               & 0.2023                & 0.2442                & 0.1906                & \multicolumn{1}{c|}{0.2597}                & 0.2243          \\
\textbf{ESMM-doQ}     & 0.1692               & 0.1598                & 0.2433                & 0.1584                & \multicolumn{1}{c|}{0.1991}                & 0.1859          \\
\textbf{ESMM-doI}     & 0.2005               & 0.2526                & 0.1815                & 0.2409                & \multicolumn{1}{c|}{0.1568}                & 0.2064          \\
\textbf{MBD}          & 0.1600               & 0.1476                & 0.2263                & 0.1471                & \multicolumn{1}{c|}{0.1848}                & 0.1731          \\ \hline
\end{tabular}}
\vspace{-10pt}
\end{table}

The results about the ablation study are presented in Table \ref{tab:ablation}. From the table, we have the following observations: 
\begin{itemize}[leftmargin=*]
    \item For the exposure fairness, the average disparities of ESMM-doI are smaller than the best baseline ESMM-RT while ESMM-doQ does not has significant improvements, indicating that $do(I)$ instead of $do(Q)$ has the main contribution to mitigate the exposure unfairness by cutting off the first backdoor path. 
    \item For the quality fairness, the average disparity of ESMM-doQ is lower than ESMM-RT, but the improvement of ESMM-doI is relatively small, indicating that $do(Q)$ does cut off the second backdoor path to effectively reduce the quality unfairness. 
    \item By combining the two intervention methods to cut off the two backdoor paths at the same time, MBD achieves the best results. Only blocking one backdoor path still suffer from the correlation along the other path, and thus only cutting off the two backdoor paths can alleviate the two unfairness simultaneity. 
\end{itemize}

%% file: 5_related.tex
\section{Related Works}
In this work, we explore how to mitigate popularity bias in multi-behavior recommendation, which is relevant to four research topics of concern: popularity bias, multiple behaviors in recommendation, causal recommendation, and fairness in recommendation. 

\vspace{-10pt}
\subsection{Popularity Bias}
Popularity bias has long been recognized as an important issue in recommender systems~\cite{abdollahpouri2017controlling,park2008long,zhu2021popularity, germano2019few, ge2020understanding, he2023exploring}. The most widely used method is IPS~\cite{IPS_rec,ips_cjl,unbiased_PS,zhang2020large}. However, there is no guaranteed accuracy for estimating propensity scores.
The other method is to eliminate the popularity bias through causal inference.
PDA~\cite{zhang2021causal} examines the deviation caused by popularity from the perspective of causality. It finds the bias amplification caused by the influence of popularity bias on item exposure. However, it does not consider the relations between popularity and other factors, such as item quality. 
\cite{xv2022neutralizing} proposes two strategies to neutralize popularity bias. The first one isolates popularity bias in one embedding direction and neutralizes the popularity direction post-training. The other one encourages all embedding directions to be disentangled and popularity neutral. $CD^{2}AN$\cite{chen2022co} disentangles item property representation and popularity representation from item property embedding and learns the comprehensive long-tail item property representation by introducing unexposed items and user behavior sequences into the item tower. CauSeR~\cite{gupta2021causer} is a session-based recommendation method, which considers popularity bias in two stages: data generation stage and training stage. It performs
deconfounded training and causal inference to remove the biases introduced during the training and simultaneously models conformity bias. However, the above causal inference methods focus on the popularity bias on one user behavior and may cause unfairness due to blindly eliminating the popularity.

\vspace{-10pt}
\subsection{Multi-behavior Recommendation}
Recommender systems learn user interests from various user behaviors and making a better recommendation based on user multiple behaviors is also a significant issue~\cite{xia2021graph, ma2018entire,xia2021knowledge}.
Unified graph can be constructed to represent multi-behavior data. In the graph, it learns behavior strength by user-item propagation layer and captures behavior semantics by item-item propagation layer~\cite{jin2020multi}.
By considering the user's click and like behavior at the same time, these works~\cite{ma2018entire,wen2020entire} alleviate the problem of data sparsity of user's like behavior through multi-task training.
There is also work that incorporates the IPS method in multi-behavior recommendation. Multi-IPW-DR~\cite{zhang2020large} considers the user's click, and like behavior at the same time. It proposes to take the probability of click as the tendency score to adjust the final post-click prediction result and use the Multi-DR method to solve the sampling deviation. However, the IPS method often has high instability because of the high variance of the propensity score.

\vspace{-10pt}
\subsection{Causal Recommendation}
Causal inference has been widely used in recommendation~\cite{he2022causpref,si2022model}.
PDA~\cite{zhang2021causal} leverages popularity by causal intervention that removes the confounding popularity bias in model training and adjusts the recommendation score with desired popularity bias in inference. However, it only focus on one user behavior and ignore the unfairness problem caused by popularity. The deconfounding technique is used in linear models to learn real interest influenced by unobserved confounders~\cite{wang2020causal} and backdoor adjustment~\cite{wang2021deconfounded} is used to alleviate bias amplification on major groups of items. 

There are works incorporating counterfactual inference in recommendation. ~\cite{wei2021model, zou2020counterfactual} uses multi-task learning to estimate the contribution of each cause and performs counterfactual inference to remove the effect of item popularity during testing. Counterfactual inference is also used to addresses the clickbait issue, which estimates the direct effect of exposure features in the prediction and removes it from recommendation scores~\cite{wang2021clicks}. 

Besides, some works learns causal embeddings to get unbiased representations for users and items. By separating user and item embeddings for interest and conformity respectively, each embedding can capture only one cause by training with cause-specific data~\cite{zheng2021disentangling}. And bias-free uniform data can be used to guide the model to learn unbiased embedding, forcing the model to discard item popularity~\cite{liu2020general}. However, obtaining such uniform data needs to randomly expose items to users, which may hurt user experience and the data is usually of a small scale which makes the learning less stable.

\vspace{-0.15cm}
\subsection{Fairness in Recommendation}
With the increasing attention on the fairness in recommendation~\cite{singh2018fairness, fu2020fairness, khademi2019fairness, kusner2017counterfactual, islam2021debiasing}, there have
been numerous attempts to define notions of fairness~\cite{mehrotra2018towards, patro2020fairrec, pitoura2020fairness, kang2020inform,wu2022joint}. Generally speaking, they define it from two perspectives: individual fairness and group fairness.
Individual fairness denotes that similar individuals (e.g., users or items) should receive similar treatments (e.g., exposure or clicks)~\cite{dwork2012fairness,xu2022regulatory}. 
Besides, group fairness indicates that decisions should be balanced for different groups distinguished by sensitive attributes like gender or race~\cite{calders2009building, zliobaite2015relation,ge2021towards}. User-oriented fairness is considered in recommendation by requiring the active and inactive user groups to be treated similarly~\cite{li2021user}. To achieve the fairness in group, adversarial learning can be introduced to enforce fairness constraints on graph embeddings, which ensures that the learned representations do not correlate with certain attributes, such as age or gender~\cite{bose2019compositional}. FairGAN~\cite{li2022fairgan} proposes a Generative Adversarial Networks (GANs) based learning algorithm mapping the exposure fairness issue to the problem of negative preferences in implicit feedback data. Counterfactual fair recommendations can also be achieved through adversary learning by generating feature-independent user embeddings for recommendation~\cite{li2021towards}. Some other methods like statistical parity-based measures ~\cite{yang2017measuring} compute the difference in the distribution of different groups for different prefixes of the ranking.~\cite{ge2021towards} proposes a reinforcement learning framework to deal with the changing group labels of items to achieve long-term fairness in recommendation.
However, these fairness-related works usually make the trade-off between ranking accuracy and fairness, and the reason possibly lies in that they blindly neglect the features of users or items.

%% file: 6_conlusion.tex
\vspace{-10pt}
\section{conclusion}
In this work, we studied how to mitigate popularity bias in multi-behavior recommendation. Firstly, we formulated two unfairness issues caused by popularity bias in multi-behavior recommendation, and inspected their underlying reasons from a causal view. Thereafter, we proposed a MBD framework to alleviate two kinds of unfairness problems, which leverages backdoor adjustment to block two backdoor paths.
We conducted extensive experiments on two real-world datasets, which justify the effectiveness of MBD on mitigating the unfairness issues and improving the accuracy.

Instead of blindly erasing the impact of popularity, this work takes the initial step to consider the effect of item quality and solve the unfairness problems caused by popularity bias in multi-behavior recommendation. In the future, there are many research directions for further exploration: 1) the analysis of popularity bias on multiple behaviors with more complex relations. For example, comment, favourite and share are three kinds of behaviors, which can be incorporated to explore their relations with popularity bias and unfairness. 2) The relations between the multiple user behaviors might be useful to reduce various biases in recommendation or information retrieval, such as selection bias and position bias. 
3) It is promising to incorporate the model-agnostic MBD framework into more recommender methods to alleviate the unfairness issues, especially under the multi-behavior recommendation settings. 


%% file: 8_acknowledgement.tex
\section*{Acknowledgment}
This work was supported by the National Natural Science Foundation of China (No.61977003) and the State Key Laboratory of Complex \& Critical Software Environment (CCSE-2024ZX-16).